\documentclass[11pt]{article}
\fontfamily{times}

\usepackage{geometry, amsmath}
\usepackage{bm}
\usepackage{bbm}
\usepackage{epsf}
\usepackage{epsfig}
\usepackage{xcolor}
\usepackage{listings}
\usepackage{amsthm}
\usepackage{mathrsfs}

\usepackage{soul}
\usepackage[style=authoryear,sorting=ynt,natbib=true,bibstyle=numeric,backend=bibtex]{biblatex}
\renewbibmacro{in:}{}
\usepackage{graphicx}
\usepackage{float}
\usepackage{caption}
\usepackage{subcaption}
\usepackage{tikz}
\usetikzlibrary{decorations.markings}
\usetikzlibrary{arrows.meta}
\usepackage{diagbox}

\newtheorem{thm}{Theorem}

\newtheorem{rem}[thm]{Remark}

\newcommand{\ben}{\begin{enumerate}}
\newcommand{\een}{\end{enumerate}}
\newcommand{\beq}{\begin{eqnarray}}
\newcommand{\eeq}{\end{eqnarray}}
\newcommand{\beqn}{\begin{eqnarray*}}
\newcommand{\eeqn}{\end{eqnarray*}}

\newcommand{\be}{\begin{equation}}
\newcommand{\ee}{\end{equation}}
\def\sk1{\vskip 10pt}

\def\uw#1{\underset {^\sim}  {#1}}

\def\b{{\bold b}}

\def\b0{{\bold 0}}

\def\bold{\bf}

\providecommand{\keywords}[1]{\textbf{\textit{Keywords: }} #1}

\def\ben{\textcolor{red}}

\usepackage{amssymb} % or amsfonts

\title{Bayesian sequential analysis of adverse events with binary data}

\author{Jiayue Wang\thanks{
Email: jwa9@iu.edu } \ and  Ben Boukai\thanks{
Email: bboukai@iu.edu}  \\ Department of Mathematical Sciences, Indiana University Indianapolis\\
Indianapolis, Indiana, 46202 }

\begin{document}
\maketitle

%\templatefigures{}

\begin{abstract}
\noindent We propose a Bayesian Sequential procedure to test hypotheses concerning the Relative Risk between two specific treatments based on the binary data obtained from the two-arm clinical trial. Our development is based on the optimal sequential test of \citet{wang2024early}, which is cast within the Bayesian framework. This approach enables us to provide, in a straightforward manner based on the Stopping Rule Principle (SRP), an assessment of the various error probabilities via posterior probabilities and conditional error probabilities. Additionally, we present the connection to the notion of the Uniformly Most Powerful Bayesian Test (UMPBT). To illustrate our procedure, we utilized the data from \citet{silva2020optimal} to analyze the results obtained from the standard Bayesian and the modified Bayesian test of \citet{berger1997unified} under several different prior distributions of the parameters involved.

\end{abstract}

\keywords{Bayesian Sequential test; Relative Risk; conditional error probabilities; Decision role.}

{\textbf{\it{AMS Classification:}}}  62L10, 62L12, 62C10

\bigskip

\setlength{\parindent}{0pt}

\section{Introduction}

Consider a clinical experiment involving a finite population of $N_0$ patients. The patients arrive sequentially and are classified according to two binary attributes; whether they were assigned to receive one of the two available treatments $A$ or $B$ and whether they exhibited a particular side effect ($Yes$ or $No$) of interest. Having classified in this manner the first $n$ patients, $(1\leq n\leq N_0)$, the results are summarized in the following $2\times 2$ table, Table \ref{t1} below. 

 \begin{table}[h]
\begin{center}{ 
\caption{Table of $n$ items with two treatment groups}
\label{t1}
\begin{tabular}{cc|c|cc|cc}
\hline
&    & &   \qquad {\it Presenting\ Side\ Effect}   &   \  \\ \hline
&  	 & 	&  $No: \ \ Y=0$ & $Yes: \ \ Y=1\ \ $  &   \  \\ \hline 
&   &	 $A$  & $n_{00}$ & $n_{01}$  &   $n_A$ \  \\ 
& {\it Treatment\  Group }  &	  &  &   &    \  \\ 
&  & $B$	& $n_{10}$ & $n_{11}$  &  $n_B:=n-n_A$ \  \\ \hline
&  Totals: & 	&  $n-S^Y_n$ &  $S^Y_n$ &  $n$ \  \\ 
\end{tabular}
}
\end{center}
\end{table}

For a give $n$, the distribution of these four counts is the multinomial distribution, so that,  
$$
{\uw n}=(n_{00}, n_{10}, n_{01}, n_{11})^\prime\sim {\mathcal MN}(n, \uw{p}), 
$$
where $\uw{p}:=(p_{00}, p_{10}, p_{01}, p_{11})^\prime$ is the vector of the corresponding probabilities, $\sum_i\sum_j p_{ij}=1$ along with $\sum_i\sum_j n_{ij}=n$. Note that the counts in Table \ref{t1} follow, marginally for a given $n$, the Binomial distribution, $n_{ij}\sim {\mathcal B}(n, p_{ij}), \  i,j=0,1$. In fact, as our sampling of units for classification is conducted sequentially, one-at-a-time, we find it useful to denote by ${\uw{z}}^{(k)}= (z_{00,k}, z_{10,k}, z_{01,k}, z_{11,k})^\prime\sim {\cal MN}(1, \uw{p})$, with $\sum_i\sum_j z_{ij, k}=1$, for each sampling stage $k=1, 2, \dots$. in which case, 
$$
{\uw{n}}=\sum_{k=1}^n{\uw{z}}^{(k)}, \ \ \ \text{and}\ \ \ n_{ij}=\sum_{k=1}^n z_{ij, k}, 
$$
In the design of the two-arm clinical trial, in which the patients (who are arriving sequentially) are assigned, {\bf at random}, to receive one of two available treatments, with some known probability $p_0$, and $q_0:= 1-p_0$, respectively, will result with $n_A=n_{00}+n_{01}$ of the patients receiving treatment $A$ and $n_B:=n-n_A$ of them receiving treatment $B$. Accordingly, for a given $n$, 
$$
n_A \sim {\cal B}(n, p_0), \ \ \ \text{with}\ \  p_0\equiv p_{00}+p_{01}, \ \ \text{($p_0$ fixed and known)},
$$
Similarly, given $n$, $n_B\sim {\cal B}(n, q_0)$. Once assigned to one of the two treatment groups, the patients are subsequently observed, during a fixed-length monitoring window, for an 'event' of interest, such as the post-treatment expression of an adverse side effect, which we generically denote $Y=1$ for 'present' and $Y=0$ for a 'absent'. We denote by $\vartheta_A$ and by $\vartheta_B$ the conditional probability for expressing (presenting) the monitored side effect given the patient received treatment $A$ or treatment $B$, respectively. Accordingly, we have
$$
\vartheta_A=\Pr( Y=1 \mid Group\ A) \quad \text{and} \ \quad \vartheta_B=\Pr(Y=1 \mid  Group\ B),  
$$ 
as well as 
$$
p_{00}=(1-\vartheta_A)\times p_0, \qquad  \text{and} \qquad p_{01}=\vartheta_A\times p_0,
$$
so that $p_{00}+p_{01}=p_0$ as desired by design. 

For such a study, one is interested in making inference concerning the unknown probabilities $\vartheta_A$ and $\vartheta_B$, and more importantly, concerning the {\bf Relative Risk} (RR) of the two treatments to express the monitored side effect, namely, $\gamma=\vartheta_A/\vartheta_B$. Note that under this two-arm clinical trial, the (total) probability for presenting the monitored side effect, which we denote by $\eta$, is therefore 
$$
Pr(Y=1)=\vartheta_A p_0+\vartheta_B q_0 =\eta. 
$$
Hence, we immediately obtain that given an observed side effect, the conditional probability it may be attributed to Treatment $A$ as, 
\be \label{1.1}
\theta(\gamma):=Pr(Group\ A \mid Y=1)=\frac{\vartheta_A p_0}{\eta} = \frac{\vartheta_A p_0}{\vartheta_A p_0+\vartheta_B q_0}\equiv \frac{1}{1+z_0/\gamma}, 
\ee
where we have put $z_0=(1-p_0)/p_0$ to indicate the {\bf{known}} matching odds of patients assigned to the treatment $B$ rather than to treatment $A$. The matching ratio $z_0$ could be thought as the ratio of the number of patients assigned to treatment $B$ relative to the number of patients assigned to treatment $A$. 

Having classified the first $n$ patients according to the above table, we denote by $S^Y_n:= n_{01}+n_{11}$ the total number of patients presenting the monitored side effect (i.e.: $Y=1$) and we denote by $X\equiv n_{01}$ the number of patients {\bf {in treatment group $A$}} that presented the monitored side effect. Clearly, given $n$, $X\sim {\cal B}(n, \ p_{01}=  \vartheta_A \times p_0)$. However, since also given $n$, $S_n^Y\sim {\cal B}(n, \eta)$ we have that given $n$ and $\{S^Y_n=m\}$, $X\sim {\cal B}(m,\ \theta(\gamma))$. \citet{silva2020optimal}, \citet{r2021exact} and \citet{silva2022bounded} refer to $m$ as the 'time' index for this two-arm binomial experiment. Accordingly, given $n$, the statistic $S^Y_n$ can be used for inference on the unknown parameter $\eta$, whereas, given $n$ and $\{S^Y_n=m\}$, the count, $X$, (out of $m$) can be used for inference on the unknown parameter $\theta(\gamma)$ and hence of the Relative Risk parameter $\gamma$.

For the 'simple' $ {\cal B}(n, \eta)$ case, \citet{wang2024early} considered the sequential test
\be \label{1.2}
H_0: \eta\leq \eta_0 \qquad \text{against}  \quad  H_1: \eta>\eta_0,
\ee
for some specified $\eta_0$, $\eta_0\in(0,1)$, which continues the sampling process as long as the null hypothesis, $H_0$, is not rejected, and stops the sampling process once there is sufficient 'evidence' to reject it. More specially, for given desired probabilities of the Type~I error, $\alpha$ and the Type~II error, $\beta$ at a given $\eta_1>\eta_0$, they determined the corresponding sample size $N_0^*$ and a critical (boundary) test value $k^*$ needed to achieve an 'optimal' fixed-sample UMP test of the hypotheses in \eqref{1.2}. Since the observations leading to the data in Table \ref{t1} become available, sequentially, one–at–a–time or in batches, the collection of the data ceases once the observed total number of patients exhibiting the monitored side effect, $S_n^Y$, exceeds the predefined threshold $k^*$, leading to the \emph{stopping time},
\be \label{1.3}
M=\inf\{n > k^*: S^Y_n>k^*\}, 
\ee
which has the Negative Binomial distribution, $M\sim {\cal NB}(k^*+1,\ \eta)$ on the integers $\{k^*+1, k^*+2, \dots \}$. Accordingly, the proposed $(\alpha, \beta)$-optimal sequential test of \eqref{1.2} may be written as:
\be \label{1.4}
\operatorname{Test_{seq}}: =
\begin{cases}
	\text{if} \ M \le N_0^* \  \ & \text{stop and reject $H_0$ in \eqref{1.2}} \\
	\text{if} \ M>N_0^* \  \ & \text{do not reject $H_0$ in \eqref{1.2}} 
\end{cases}
.
\ee
Note that, in either case, the final number of observations, at 'termination', is $M^*=\min\{M, N_0^*\}$, which is the total number of observations collected and used to arrive at a conclusion of the test for \eqref{1.2}. Note, however, that unlike the Sequential Probability Ratio Test (SPRT), the sequential collection of the data under the stopping rule \eqref{1.3} is continued as long as $H_0$ is not rejected. Accordingly, upon 'termination' (either by reaching the maximal number of observations $N_0^*$ or by rejecting $H_0$ beforehand), $S^Y_{M^*}\in \{0, 1, 2, \dots, k^*+1\}$. 

\citet{kulldorff2011maximized}, \citet{silva2020optimal} have considered the above two-arm binomial experiment and the Relative Risk (RR), $\gamma=\vartheta_A/\vartheta_B$, of the two treatments to present the monitored side effect. Thus they considered the testing problem of the hypotheses concerning the parameter $\gamma$, in the form
\be \label{1.5}
H_0: \gamma=1 \qquad \text{against} \qquad \ H_1:  \gamma> 1, 
\ee
which are equivalent to the hypotheses $
H_0: \vartheta_A=\vartheta _B$ against $H_1: \vartheta_A>\vartheta_B$. They proposed an elaborate alpha-spending approach for calculating the 'stopping-rule' for the sequential test of the hypotheses in \eqref{1.5}, concerning the Relative Risk, $\gamma$. \citet{r2021exact} presented a linear programming framework to determine optimal alpha spending that minimizes expected time to signal, or expected sample size as needed, leading to bound the width of the confidence interval for $\gamma$ at the terminal analysts.

For a specified matching allocation ratio between the control group and the treatment group, $z_0$, we denote $\theta_0:=(1+z_0)^{-1}$. We note that the hypotheses in \eqref{1.5} can be stated equivalently in terms of $\theta\equiv \theta(\gamma)\equiv \gamma/(z_0+\gamma)$ as 
\be \label{1.6}
H_0:\theta =\theta_0 \ \ \ \text{(i.e. with $\gamma=1$)} \quad \text{against} \quad
H_1:\theta>\theta_0\ \ \ \text{(i.e. with $\gamma>1$)}.
\ee
For instance, if the matching allocation ratio is $1:1$ (balanced), then $z_0=1$ and therefore $\theta_0=1/2$, leading to the test of 
$$
H_0: \theta=1/2 \quad\text{against} \quad  H_1: \theta > 1/2.
$$
Hence, the optimal sequential test of \citet{wang2024early} for hypotheses in the form of \eqref{1.2} can be applied in this current situation too.

In Section \ref{s2}, we consider the testing problem of general hypotheses in the form of \eqref{1.6} and related inference problems to the Relative Risk, $\gamma$, entirely within the Bayesian framework. Instead of determining a value of $\theta$ under $H_1$ to calculate the sequential test statistics, the Bayesian sequential test uses the entire alternative support to calculate the corresponding Bayes Factor. The resulting sequential Bayesian test does not necessitate any alpha-spending calculations (of the type provided by \citet{silva2020optimal}). Additionally, we present a modified Bayesian test (as was advocated in \citet{berger1997unified}) which benefits from the easily calculated conditional error probabilities and has both the classical (conditional) frequentist as well as Bayesian interpretation.

In Section \ref{s3}, we apply our Bayesian sequential test and the modified Bayesian sequential test to the data discussed in \citet{silva2020optimal} and \citet{silva2022bounded} under different alternatives as well as different priors. Specifically, based on the modified Bayesian test, we calculate the conditional Type~I or Type~II error probability for each data point with corresponding decision we make. Furthermore, we connect our results to the Uniformly Most Powerful Test (UMPBT) which can be compared from the frequentist perspective. In Section $4$, we compare our results with the results presented in \citet{silva2020optimal} and \citet{silva2022bounded} and provide some closing remarks.

\section{The Bayesian framework for Binomial Probabilities} \label{s2}

\subsection{The Standard Bayesian Test}

Suppose that for a given $m$ and unknown $\theta\in (0,1)\equiv \Theta$, a binomial random variable, $X_m\sim{\cal B}(m, \theta)$ with a $pmf$ given by 
$$
p(x \mid \theta, \ m)=Pr(X_m=x \mid \theta,\ m)=\genfrac{(}{)}{0pt}{}{m}{x}\theta^{x}(1-\theta)^{m-x}\mathbf{I}[x=0, 1, \dots, m].
$$
We will see below that the structure of the hypotheses in \eqref{1.2} can generally be formulated as 
\be \label{2.7}
H_0: \theta\in  \Theta_0 \qquad \text{against}\qquad  H_{1}: \theta\in \Theta_1  
\ee
for some $\Theta_0\subset \Theta$ and $\Theta_1=\Theta- \Theta_0$. We assume that given that the hypothesis $H_i$ is true, $\theta$ has a proper prior $pdf$, $p_i(\cdot )$ over $\Theta_i$, $i=0,1$, so that 
$$
\int_{\Theta_i} p_i(\theta)d\theta =1, \qquad \qquad i=0,1.
$$
Accordingly, the test of the hypotheses in \eqref{2.7} can equivalently be formulated as the simple hypotheses concerning the marginal distribution of $X_m$, namely,
\be\label{2.8}
H_0: X_m\sim f_{0, m}(x)  \qquad \text{against}\qquad  H_{1}: X_m\sim f_{1, m}(x),    
\ee
where for a given $m$ and $i=0,1$, $f_{i,m} (\cdot)$ is the marginal distribution of $X_m$ under $H_i$, given by
$$
f_{i, m}(x):=\int_{\Theta_i}p(x \mid \theta,  m) \times p_i(\theta) d\theta, 
$$
respectively. In the Bayesian framework, one typically assumes that the prior probability that $H_0$ is true, is $Pr(H_0)=\pi_0$ for some $0<\pi_0<1$, and then proceeds to obtain the posterior probability of $H_0$ being true, {\bf{given}} the data $\{X_m=x\}$, namely $Pr(H_0 \mid X_m= x)$, where, in the case of the hypotheses \eqref{2.8}, 
\be \label{2.9}
Pr(H_0 \mid X_m= x)=\frac{f_{0, m}(x)\pi_0}{f_{0, m}(x)\pi_0+f_{1, m}(x)(1-\pi_0)}=\frac{\ell\times B_m(x)}{1+\ell \times B_m(x)},
\ee
where we have substituted $\ell:= \pi_0/(1-\pi_0)$ and $B_m(x):=f_{0,m}(x)/f_{1,m}(x)$. The ratio $B_m(x)$, is the so-called Bayes factor of $H_0$ to $H_1$ (see, for example, \citet{berger1997unified}). Note that when $\pi_0=1/2$, so that $\ell\equiv 1$, which indicates that the prior probability of the null hypothesis is equal to the prior probability of the alternative hypothesis, the posterior probability of $H_0$ being true, given the data $\{X_m=x\}$ simplifies to
$$
Pr(H_0 \mid X_m=x)=\frac{B_m(x)}{1+ B_m(x)}. 
$$
Hence, in this basic fixed-sample setup, the Bayesian would simply reject $H_0$ in favor of $H_1$ if $\Pr(H_0 \mid X_m=x)<\pi_0\equiv  1/2$ or equivalently, if $B_m(x)<1$. That is, for a given $m$ and $\ell=1$, the Bayesian (fixed-sample) test is,
\be \label{2.10}
\operatorname{Test_{Bayes}}: =
\begin{cases}
	\text{if} \ B_m(x)<1 \  \ & \text{Reject $H_0$ in \eqref{2.8}} \\
	\text{if} \ B_m(x) \ge 1 \  \ & \text{Accept  $H_0$ in \eqref{2.8}} .
\end{cases}
\ee 
However, the Bayesian test in \eqref{2.10} might lead to certain ambiguities, especially whenever the value of the Bayes factor $B_m(x)$ is close to the boundary of $1$. Based on \citet{jeffreys1961edition}, \citet{kass1995bayes} suggestions of varying levels of 'rejection of $H_0$' all according to the amount of evidence in the data measured by the index $J_B=\log_{10}(B_m(x))$ are presented in Table \ref{t2}.

\begin{table}[H]
\begin{center}{\small 
\caption{The values of Bayes factor $B_m$ as a summary of the evidence provided by the data.}
\label{t2}
\begin{tabular}{ c c c }
\hline
$J_B$ & $B_m$ & Evidence against $H_0$ \\
\hline
$-1/2$ to $0$ & $0.3162$ to $1$ & Not worth more than a bare mention \\
$-1$ to $-1/2$ & $0.1$ to $0.3162$ & Substantial \\
$-2$ to $-1$ & $0.01$ to $0.1$ & Strong \\
$<-2$ & $<0.01$ & Decisive \\
\hline
\end{tabular}
}
\end{center}
\end{table} 

Accordingly, one would substantially reject $H_0$ if $-1 < J_B < -1/2$, and would strongly reject $H_0$ if $-2< J_B <-1$ and decisively reject $H_0$ if $J_B<-2$.

As is apparent from \eqref{2.9}, the calculations of the Bayes Factor require specification of the prior distribution for $\theta$. For the binomial model in \eqref{2.7}, a natural choice of prior distributions for $\theta$ would be the conjugate class of distributions over $\Theta=(0,1)$, so that $\theta\sim {\cal B}eta(a, b)$ for some parameters $a>0$, $b>0$. Hence, the prior $pdf$ of $\theta$ is given by 
$$
p(\theta)=\frac{\theta^{a-1}(1-\theta)^{b-1}}{ beta(a, b)}\mathbf{I}[0 < \theta < 1],  
$$
where $beta(a, b):=\mathcal{I}_1(a, b)$, and where for any $\xi \in [0,1]$, 
$$
\mathcal{I}_\xi(a, b)=\int_0^\xi u^{a-1}(1-u)^{b-1}du,
$$
is the incomplete beta function. With this (conjugate) Beta prior distribution for $\theta$ over $\Theta=(0,1)$, the posterior distribution of $\theta$, given the data $X_m=x$, is the ${\mathcal B}eta(x+a, m-x+b)$ distribution, so that given $m$ and $X_m=x$, 
\be \label{2.11}
p(\theta \mid X_m)= \frac{\theta^{x+a-1}(1-\theta)^{m-x+b-1}}{ beta(x+a, m-x+b)}\mathbf{I}[0 < \theta < 1].  
\ee
Further, in this case, the marginal probability distribution of $X_m$ is the Beta-Binomial distribution, given by 
$$
f_{m}(x) = \Pr(X_m=x)= \genfrac{(}{)}{0pt}{}{m}{x} \frac{beta (a+x, m-x+b)}{beta(a, b)}\mathbf{I}[x=0, 1, \dots, m]. 
$$
As an immediate application, consider the binomial experiment as summarized in Table \ref{t1} above, in which the sequential collection of the data was terminated after $n$ observations according to the stopping rule in \eqref{1.3}, so that for the given $n$ and $\{S_n^Y=m\}$, $X_m\sim {\mathcal B}(m, \theta(\gamma))$, where, as in \eqref{1.6}, $\theta\equiv \theta(\gamma)=\gamma/(z_0+\gamma)$ with a known values $z_0$. We remind the reader that according to the {\it Stopping Rule Principle} (SRP), posterior probabilities, as evidentiary data measures, remain unaffected by the optional stopping rule for the data collection, see \citet[p.~74]{berger1988likelihood} or \citet[p.~502]{berger2013statistical}. Hence, in this case too, with the stopping time $M$ in \eqref{1.3}, the expression \eqref{2.9} for the posterior probability of $H_0$ holds true also given the event ${\mathcal A}_{nm}:= \{(M=n)  \  \text{\ {and}}\  (S^Y_{n}=m) \}$ for some $n\geq m$ and $m\in \{0, 1, 2, \dots \}$. That is 
$$
Pr(H_0 \mid X_m=x, {\cal A}_{nm})\equiv Pr(H_0 \mid X_m=x)=\frac{\ell\times B_m(x)}{1+\ell \times B_m(x)}.
$$
In similarity to the hypotheses in \eqref{1.6}, we consider two additional cases of hypotheses concerning the RR parameter $\gamma$. In all cases, we will assume equal prior probabilities of the null and the alternative hypotheses, so that, in all cases we take $\pi_0=1/2$ (so that $\ell=1$) in \eqref{2.9}.

\noindent {\bf Case 1:} We are interested in testing of $H_0: \gamma=1$ against $H_1:\gamma \neq 1$, which with $\theta_0=(1+z_0)^{-1}$, can readily be seen as equivalent to the testing of 
\be \label{2.12}
H^{(1)}_0: \theta=\theta_0 \qquad \text{against} \qquad H^{(1)}_1:\theta\neq \theta_0. 
\ee
In this case, for $x=0, 1, \dots, m$, the marginal distributions of $X_m$ under $H_0^{(1)}$ and $H_1^{(1)}$, are 
$$
f^{(1)}_{0, m}(x)= \genfrac{(}{)}{0pt}{}{m}{x}\theta_0^{x}(1-\theta_0)^{m-x}
$$ 
and 
$$
f^{(1)}_{1,m}(x)= \genfrac{(}{)}{0pt}{}{m}{x}\frac{beta(a+x, m-x+b)}{beta(a,b)}, 
$$
respectively. Hence the Bayes factor of $H^{(1)}_0$ to $H^{(1)}_1$, in \eqref{2.12} is 
\be \label{2.13}
B^{(1)}_m(x)\:= \frac{f^{(1)}_{0,m}(x)}{f^{(1)}_{1,m}(x)}= \frac{ beta(a, b)\times \theta_0^{x}(1-\theta_0)^{m-x}}{beta(a+x, m-x+b)}.
\ee
Accordingly, upon termination of the observation process at $\{M=n\}$ for some $n$, with a given total number of observed side effects $\{S_n^Y=m\}$, of which $X_m=x$ of them are 'attributed' to treatment $A$, we may calculate the value of the Bayes Factor $B^{(1)}_m(x)$, in \eqref{2.13}, for any choice of $a, b$ and $z_0$. Therefore, we can obtain the corresponding posterior probability of $H^{(1)}_0$ being true (given the data $\{X_m=x\}$), as given in \eqref{2.18} below.

\noindent {\bf Case 2:} In a similar fashion, the testing of $H_0: \gamma = 1$ against $H_1:\gamma > 1$, which with $\theta_0=(1+z_0)^{-1}$, can readily be seen as equivalent to the testing of 
\be \label{2.14}
H^{(2)}_0: \theta = \theta_0 \qquad \text{against} \qquad H^{(2)}_1:\theta> \theta_0, 
\ee
In this case too, with equal prior probabilities for $H^{(2)}_0$ and $H^{(2)}_1$, (i.e., $\pi_0=1/2$), we have for $x=0, 1, \dots, m$, 
$$
f^{(2)}_{0, m}(x)= \genfrac{(}{)}{0pt}{}{m}{x}\theta_0^{x}(1-\theta_0)^{m-x}
$$ 
and 
$$
f^{(2)}_{1,m}(x)= \genfrac{(}{)}{0pt}{}{m}{x}\frac{beta(a+x, m-x+b)- \mathcal{I}_{\theta_0}(a+x, m-x+b)}{beta(a,b)}.
$$
Hence the Bayes factor of $H^{(2)}_0$ to $H^{(2)}_1$, in \eqref{2.14} is 
\be \label{2.15}
B^{(2)}_m(x)\:= \frac{f^{(2)}_{0,m}(x)}{f^{(2)}_{1,m}(x)}= \frac{ beta(a, b)\times \theta_0^{x}(1-\theta_0)^{m-x}}{{beta(a+x, m-x+b)- \mathcal{I}_{\theta_0}(a+x, m-x+b)}}.
\ee
Here too, upon termination of the observation process at $\{M=n\}$ for some $n$, with a given total number of observed side effects $\{S_n^Y=m\}$, of which $X_m=x$ of them are 'attributed' to treatment $A$, we may calculate the value of the Bayes Factor $B^{(2)}_m(x)$, in \eqref{2.15}, for any choice of $a$, $b$ and $z_0$, and therefore obtain the corresponding posterior probability of $H^{(2)}_0$ being true (given the data $\{X_m=x\}$), as given in \eqref{2.18} below.  

\noindent {\bf Case 3:} Here we are testing $H_0: \gamma \leq 1$ against $H_1:\gamma > 1$, which is equivalent to testing of 
\be \label{2.16}
H^{(3)}_0: \theta\leq \theta_0 \qquad \text{against} \qquad H^{(3)}_1:\theta> \theta_0, 
\ee
with $\theta_0=(1+z_0)^{-1}$. In this case too, with equal prior probabilities for $H^{(3)}_0$ and $H^{(3)}_1$, (i.e., $\pi_0=1/2$), we have for $x=0, 1, \dots, m$, 
$$
f^{(3)}_{0, m}(x)= \genfrac{(}{)}{0pt}{}{m}{x}\frac{\mathcal{I}_{\theta_0}(a+x, m-x+b)}{beta(a,b)}
$$
and 
$$
f^{(3)}_{1,m}(x)= \genfrac{(}{)}{0pt}{}{m}{x}\frac{beta(a+x, m-x+b)- \mathcal{I}_{\theta_0}(a+x, m-x+b)}{beta(a,b)}.
$$
Hence the Bayes factor of $H^{(3)}_0$ to $H^{(3)}_1$, in \eqref{2.16} is 
\be \label{2.17}
B^{(3)}_m(x)\:= \frac{f^{(3)}_{0,m}(x)}{f^{(3)}_{1,m}(x)}= \frac{\mathcal{I}_{\theta_0}(a+x, m-x+b)}{{beta(a+x, m-x+b)- \mathcal{I}_{\theta_0}(a+x, m-x+b)}}.
\ee
Clearly, the calculation of the posterior probability is straightforward as in the previous cases. In fact, in all Cases 1-3, we reject $H^{(i)}_0$ if $B^{(i)}_m<1$, and report the respective posterior probabilities of $H^{(i)}_0$ and $H^{(i)}_1$, being true as 
\be \label{2.18}
\alpha^*\left(B^{(i)}_m(x) \right)=\Pr \left(H_0^{(i)} \mid X_m=x \right)=\frac{B^{(i)}_m(x)}{1+B^{(i)}_m(x)},\ \quad \text{for}\ \  i=1, 2, 3;
\ee
\be \label{2.19}
\beta^*\left(B^{(i)}_m(x) \right)=\Pr \left(H_1^{(i)} \mid  X_m=x \right)=\frac{1}{1+B^{(i)}_m(x)},\ \quad \text{for}\ \  i=1, 2, 3.
\ee
Accordingly, in all these three cases, the Bayesian test can be presented as,
\be \label{2.20}
\operatorname{Test_{Bayes}}: =
\begin{cases}
	\text{if} \ B^{(i)}_m(x)<1, \  \  \ & \text{Reject $H^{(i)}_0$ and report the} \\
 & \text{posterior probability $\alpha^*\left(B^{(i)}_m(x) \right)$}\\
	\text{if} \ B^{(i)}_m(x) \ge 1, \ \  \ & \text{Accept $H^{(i)}_0$ and report the}\\
 & \text{posterior probability $\beta^*\left(B^{(i)}_m(x) \right)$},
\end{cases}
\ee
for $i=1,2,3$. Clearly, the Bayesian tests in \eqref{2.20} can incorporate Jeffrey's criteria as in Table \ref{t2} to determine the strength of the data in evidence to support the rejection of the respective $H_0^{(i)}$.

Note that to fully apply the test of the hypotheses in \eqref{2.8}, one would need to specify the values of the prior distribution parameters $a$ and $b$ and matching allocation ratio between the two treatment groups, $z_0$, in the calculations of $B^{(i)}_m(x)$, $i=1, 2, 3$.

\subsection{The Modified Bayesian Test}

In the spirit of Table \ref{t2}, we modify the standard Bayesian test to include a 'no-decision region'. This would often be a useful approach in the situation when there is an ambiguous result showing up. 

As advocated and suggested in \citet{berger1997unified}, we modify the Bayesian test in \eqref{2.8} to include a no-decision region in it. The modified Bayesian test has the form, 
\be \label{2.21}
\operatorname{Test_{M-Bayes}}: =
\begin{cases}
	\text{if} \ B_m(x)< r \  \ & \text{Reject $H_0$ and report the conditional} \\
 & \text{error probability $\alpha^*\left(B_m(x) \right)$} \\
 	\text{if} \  r \le  B_m(x) \le a \  \ & \text{make no decision} \\
	\text{if} \ B_m(x)>a \  \ & \text{Accept $H_0$ and report the conditional}\\
 & \text{error probability $\beta^*\left(B_m(x) \right)$},
\end{cases}
\ee
where $r<a$ are two constants, defined and calculated as described in \eqref{2.22} below.

Let $F_0$ and $F_1$ be the {\it{cdf}} of $B_m(x)$ under $H_0$ and $H_1$, respectively. Their inverse $F_0^{-1}$ and $F_1^{-1}$ exist over the range $\mathscr{B}$ of $B_m(x)$. For any $b \in \mathscr{B}$, we let
$$
\psi(b):=F_0^{-1} \left(1-F_1(b)\right) \quad \text{and} \quad \psi^{-1}(b):=F_1^{-1} \left(1-F_1(b)\right).
$$
Accordingly, the decision constants $r$ and $a$ are calculated as, 
\be \label{2.22}
\begin{split}
&r=1 \quad \text{and}  \quad a=\psi(1) \quad \text{if} \ \psi(1) \ge 1, \\
&r=\psi^{-1}(1)  \quad \text{and}  \quad a=1 \quad \text{if} \ \psi(1) < 1.
\end{split}
\ee

As stated in \citet{berger1997unified}, the posterior probabilities of $H_0$ and $H_1$ could also be interpreted as the conditional error frequentist Type~I and Type~II probabilities. $\operatorname{Test_{M-Bayes}}$ is also a conditional frequentist test, arising from the use of the conditioning statistic
\be \label{2.23}
S(X)=\min\{B_m(X),\psi^{-1}(B_m(X))\},
\ee
over the domain $\mathscr{X}^*=\{0<S(X)<r\}$ (the complement of $\mathscr{X}^*$ is the no-decision region). Therefore, based on \eqref{2.18} and \eqref{2.19}, the conditional error probabilities upon accepting or rejecting which are in complete agreement with the Bayesian posterior probabilities are
\be \label{2.24}
\alpha^*(s)=\frac{s}{1+s} \quad \text{and} \quad \beta^*(s)=\frac{1}{1+\psi(s)}.
\ee

\section{Example} \label{s3}

\citet{silva2020optimal} reported the results of a sequential study on the side effects of the H1N1 influenza vaccine during the 2009–2010 influenza season (see \citet{lee2011h1n1}) which was conducted with equal allocation between the exposed group, $A$, and the not exposed group, $B$, so that study design matching constant is $z_0=1$ and hence, $\theta_0=1/2$. The 24 successive data points on $(m,\ X_m)$ from the study cited there are presented in the table below along with the values of $\hat\gamma \equiv X_m/(m-X_m)$. Applying their iterative 'sequential' procedure to the hypotheses of Case $2$, \citet{silva2020optimal} have determined for $\alpha=0.05$ that the testing procedure should have been terminated at the $19^{th}$-group data point, with $m=222$ and $X_{222}=134$, which resulted with estimated RR of $\hat \gamma=1.5227$. For Case $3$ and a related confidence interval, \citet{silva2022bounded} have determined, for $\alpha=0.1$, that the testing procedure should have been terminated at the $18^{th}$-group data point, with $m=218$ and $X_{218}=130$, which resulted with an estimated RR of $\hat \gamma=1.4773$.   

For comparison, we apply the above Bayesian Beta-Binomial models we propose to these 24 data points under several choices of prior distributions, however, all with prior mean of $0.5$.

\subsection{The Standard Bayesian Test} \label{s3.1}

\noindent I) {\it{Non-informative (uniform) prior distribution}}

We assume a uniform prior distribution for $\theta$, so that $\theta\sim {\mathcal B}eta (a=1, b=1)$. We calculated the Bayes factor $B^{(i)}_m$ and the corresponding posterior probability $Pr(H_0^{(i)} \mid X_m=x)$, $i=1, 2, 3$, for the hypotheses tests on the RR, $\gamma$, in the three cases we considered above. The numerical results are provided in Table \ref{t3} below.

\begin{figure}[H]
\centering
\begin{subfigure}[b]{0.328\textwidth}
\centering
\includegraphics[width=\textwidth]{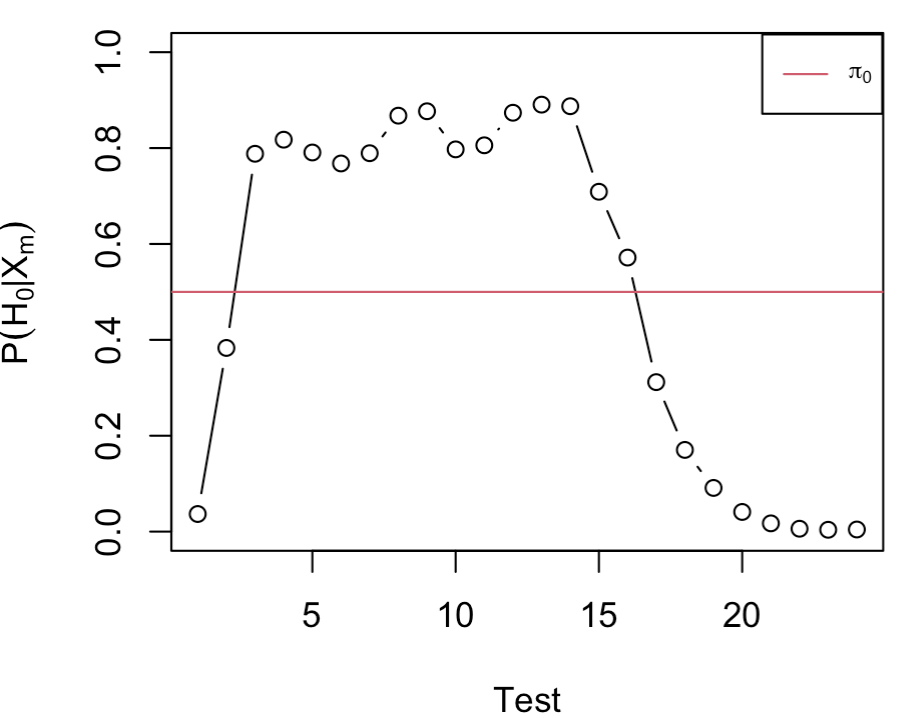}
\caption{Case 1}
\end{subfigure}
\hfill
\begin{subfigure}[b]{0.328\textwidth}
\centering
\includegraphics[width=\textwidth]{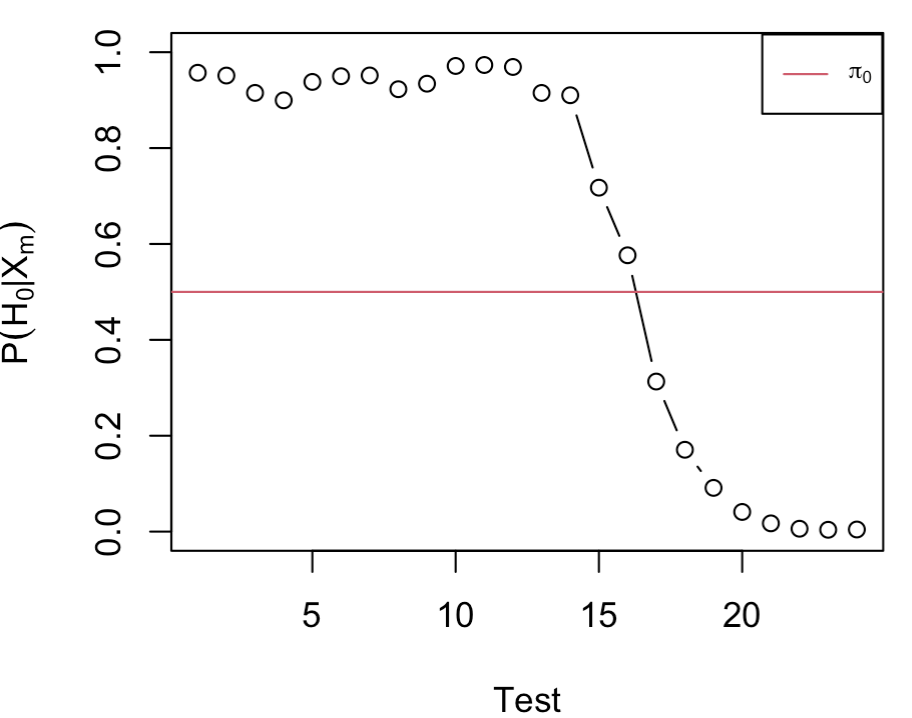}
\caption{Case 2}
\end{subfigure}
\begin{subfigure}[b]{0.328\textwidth}
\centering
\includegraphics[width=\textwidth]{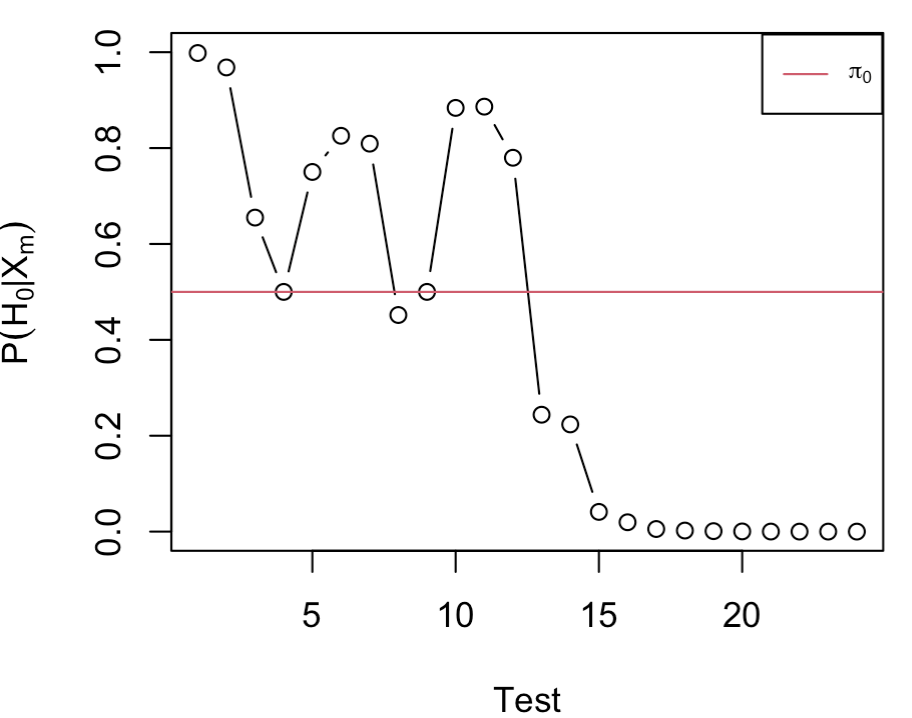}
\caption{Case 3}
\end{subfigure}
\caption{The posterior probability of $H_0$ for the 24 data points of Table \ref{t3}}
\label{f1}
\end{figure}

As can be seen, for the hypotheses of Case $1$ and Case $2$, our Bayesian model above has led to the potentially earlier 'termination' at the $18^{th}$-group data point with a total of $m=218$ 'events' of observed side effect of which $X_m=130$ were noted in the treatment group with an estimated relative risk of $\hat \gamma=X_m/(m-X_m)=1.4773$. The resulting Bayes Factors are $B_{218}^{(1)}=0.2055$ and $B_{218}^{(2)}=0.2059$, which are less than $0.3162$ based on the standard shown in Table \ref{t2}. This leads to a rejection of $H_0^{(i)}$, $i=1, 2$ and the posterior probabilities of the null hypothesis and the alternative hypothesis given the data are
$$
\Pr(H_0^{(1)}: \gamma=1 \mid X_{218}=130)=0.1704,  \qquad \Pr(H_1^{(1)} :\gamma \neq 1 \mid X_{218}=130)=0.8296,
$$
and
$$
\Pr(H_0^{(2)} :\gamma= 1 \mid X_{218}=130)=0.1707, \qquad \Pr(H_1^{(2)} :\gamma > 1 \mid X_{218}=130)=0.8293,
$$
which are clearly smaller than the assumed prior probability $\pi_0=0.5$ of $H_0^{(i)}$, $i=1,2$, being true (marked in a red line in Figure \ref{f1}).

On the other hand, for the hypotheses of Case $3$, one would 'terminate' the observation process, much earlier, already at the $14^{th}$-group data point with a total of $m=172$ 'events' of observed side effect of which $X_m=91$ were noted in the treatment group, leading to the estimated relative risk of $\hat \gamma=1.1235$. The resulting Bayes Factor is $B_{172}^{(3)}=0.2880$ which are less than $0.3162$ based on the standard shown in Table \ref{t2} and the posterior probabilities of the null hypothesis and the alternative hypothesis given the data are 
$$
\Pr(H_0^{(3)} :\gamma \le 1 \mid X_{172}=91)=0.2236,  \quad \Pr(H_1^{(3)} :\gamma > 1 \mid X_{172}=91)=0.7764,
$$
which are clearly smaller than the assumed prior probability $\pi_0=0.5$ of $H_0^{(3)}$ being true (marked in a red line in Figure \ref{f1}). 

\begin{rem} \label{r1}
We point out that in Case $3$, we choose the early 'termination' at the $14^{th}$-group data point instead of $8^{th}$-group data point. At the $8^{th}$-group data point with a total of $m=67$ 'events' of observed side effect of which $X_m=34$ were noted in the treatment group, leading to the estimated relative risk of $\hat \gamma=X_m/(m-X_m)=1.0303$. The resulting Bayes Factor is $B_{67}^{(3)}=0.8241<1$, and the posterior probabilities of the null hypothesis and the alternative hypothesis given the data are 
$$
\Pr(H_0^{(3)} :\gamma \le 1 \mid X_{67}=34)=0.4518, \quad \Pr(H_1^{(3)} :\gamma > 1 \mid X_{67}=34)=0.5482.
$$
Although the value of $\Pr(H_0^{(3)}\mid X_{67}=34)$ is less than $0.5$, following the standard shown in Table \ref{t2}, the evidence against $H_0^{(3)}$ at the $8^{th}$-group data point {\bf does not worth more than a bare mention}. Therefore, we would substantially reject $H_0^{(3)}$ and stop at the $13^{th}$-group data point.
\end{rem}

\begin{rem} \label{r2}
In Case $3$, we observe that the values of Bayes Factor at the $4^{th}$ and $9^{th}$-group data point are exactly equal to $1$. Due to the data structure, we have the matching allocation ratio $z_0=1$ and at the $4^{th}$-group and $9^{th}$-group data points, $m-X_m=X_m$. Moreover, since the mean of the chosen prior distributions is $0.5$, it is not surprising that at these two specific data points, the values of Bayes Factor are $1$.
\end{rem}

\begin{table}[H]
\begin{center}{\footnotesize
\caption{The values of Bayes factor and posterior probability of $H_0$ for each data point among the 3 different cases with the uniform prior.}
\label{t3}
\begin{tabular}{ c c c c c c c c c c }
\hline
Test & $m$ & $X_m$ & $\hat{\gamma}$ & $B_m^{(1)}$ & $\Pr(H_0^{(1)}\mid X_m)$ & $B_m^{(2)}$ & $\Pr(H_0^{(2)}\mid X_m)$ & $B_m^{(3)}$ & $\Pr(H_0^{(3)}\mid X_m)$ \\
\hline
1  & 12  & 1 &0.0909 &0.0381 &0.0367 &22.2857 &0.9571 &584.1429 &0.9983\\
2   &18   &5 &0.3846 &0.6210 &0.3831 &19.5382 &0.9513  &30.4623 &0.9682\\
3   &24  &11 &0.8462 &3.7195 &0.7881 &10.7807 &0.9151   &1.8984 &0.6550\\
4   &30  &15 &1.0000 &4.4784 &0.8175  &8.9568 &0.8996   &1.0000 &0.5000\\
5   &34  &15 &0.7895 &3.7811 &0.7908 &15.1377 &0.9380   &3.0035 &0.7502\\
6   &40  &17 &0.7391 &3.3088 &0.7679 &18.9675 &0.9499   &4.7325 &0.8256\\
7   &46  &20 &0.7692 &3.7458 &0.7893 &19.6273 &0.9515   &4.2398 &0.8092\\
8   &67  &34 &1.0303 &6.5554 &0.8676 &11.9580 &0.9228   &0.8241 &0.4518\\
9   &78 &39 &1.0000 &7.1142 &0.8768 &14.2285 &0.9343    &1.0000 &0.5000\\
10 &100  &44 &0.7857 &3.9342 &0.7973 &33.8717 &0.9713   &7.6095 &0.8838\\
11 &115  &51 &0.7969 &4.1504 &0.8058 &36.5229 &0.9733   &7.7999 &0.8864\\
12 &135  &63 &0.8750 &6.9197 &0.8737 &31.4257 &0.9692   &3.5415 &0.7798\\
13 &167  &88 &1.1139 &8.1376 &0.8906 &10.7609 &0.9150   &0.3224 &0.2438\\
14 &172  &91 &1.1235 &7.8705 &0.8873 &10.1371 &0.9102   &0.2880 &0.2236\\
15 &190 &107 &1.2892 &2.4347 &0.7089  &2.5390 &0.7174   &0.0429 &0.0411\\
16 &197 &113 &1.3452 &1.3341 &0.5716  &1.3606 &0.5764   &0.0199 &0.0195\\
17 &211 &124 &1.4253 &0.4531 &0.3118  &0.4556 &0.3130   &0.0055 &0.0054\\
18 &218 &130 &1.4773 &0.2055 &0.1704  &0.2059 &0.1707   &0.0022 &0.0022\\
19 &222 &134 &1.5227 &0.1003 &0.0911  &0.1004 &0.0912   &0.0010 &0.0010\\
20 &231 &141 &1.5667 &0.0427 &0.0410  &0.0427 &0.0410   &0.0004 &0.0004\\
21 &240 &148 &1.6087 &0.0175 &0.0172  &0.0175 &0.0172   &0.0001 &0.0001\\
22 &245 &153 &1.6630 &0.0060 &0.0060  &0.0060 &0.0060   &0.0000 &0.0000\\
23 &247 &155 &1.6848 &0.0039 &0.0038  &0.0039 &0.0038   &0.0000 &0.0000\\
24 &251 &157 &1.6702 &0.0044 &0.0044  &0.0044 &0.0044   &0.0000 &0.0000\\
\hline
\end{tabular}
}
\end{center}
\end{table} 

We further applied Jeffrey's prior discussed in \citet{jeffreys1946invariant}, as a 'objective' and non-informative prior to above three cases. Specifically, for binomial model, the Jeffrey's prior is $\mathcal{B}eta (1/2,1/2)$. Since the results under the Jeffrey's prior are similar to the results we obtained  under the uniform prior, we would not give detailed interpretation of the results under the Jeffrey's prior (shown in Appendix).

\noindent II) {\it{Informative  prior distribution}}

Here we assume an informative prior distribution for $\theta$ that is induced by determining the values of $(a,b)$ by the following set of two equations, 
$$
E(\theta)=\frac{1}{1+z_0} \quad \text{and}  \quad \Pr(\lvert \gamma (\theta)-1 \rvert \le \epsilon \mid \theta )=\delta
$$ 
for some given $\delta$ and $\epsilon$, and where $\gamma(\theta)\equiv  z_0 \times \theta/(1-\theta)$ by \eqref{1.1}. In the current  example, we have $z_0=1$ and assume $\epsilon=0.1$, and $\delta=0.55$. Hence, we 'solve' for our prior parameters to obtain that $a=b=113.8288$. Applying the procedure we proposed above, we calculated the Bayes factor $B^{(i)}_m$ and the corresponding posterior probability $\Pr(H_0^{(i)} \mid X_m=x)$, $i=1,2,3$, for the hypotheses testing on $\gamma$ in the three cases we considered above. The numerical results are provided in Table \ref{t4} below.

\begin{figure}[H]
\centering
\begin{subfigure}[b]{0.328\textwidth}
\centering
\includegraphics[width=\textwidth]{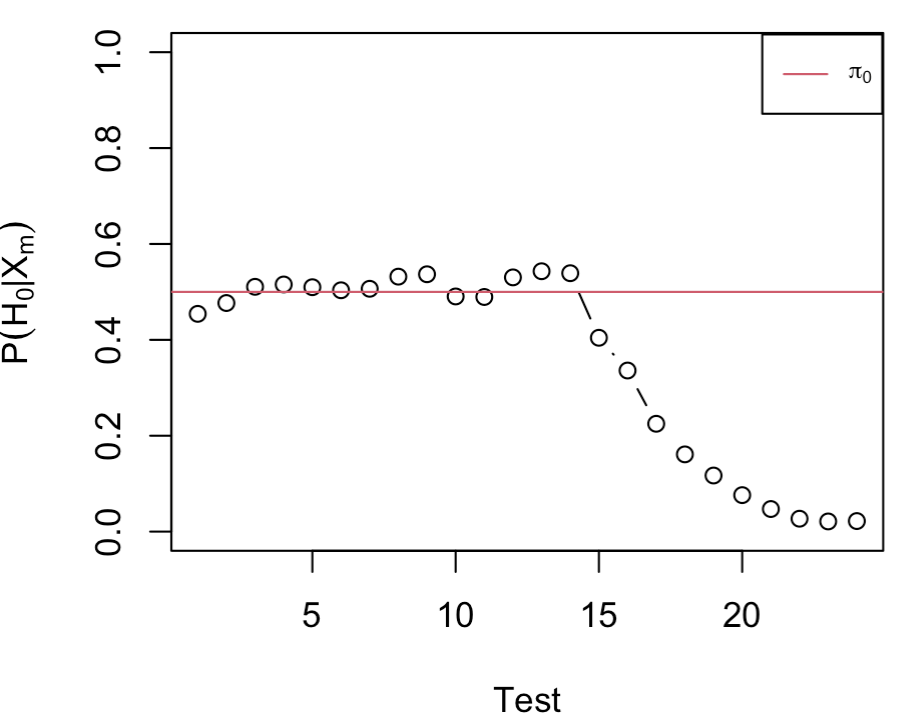}
\caption{Case 1}
\end{subfigure}
\hfill
\begin{subfigure}[b]{0.328\textwidth}
\centering
\includegraphics[width=\textwidth]{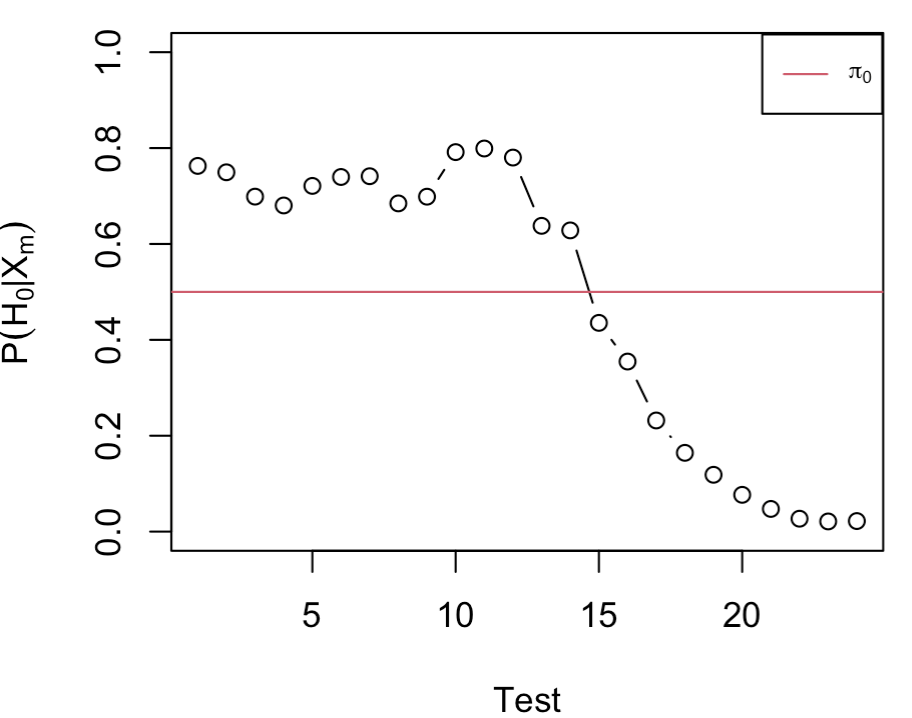}
\caption{Case 2}
\end{subfigure}
\begin{subfigure}[b]{0.328\textwidth}
\centering
\includegraphics[width=\textwidth]{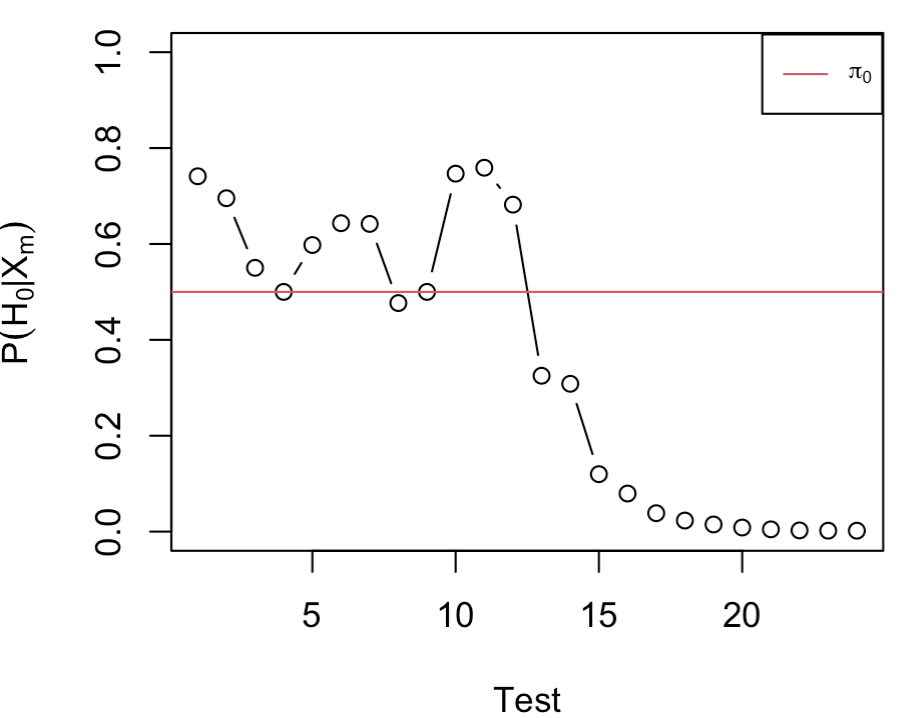}
\caption{Case 3}
\end{subfigure}
\caption{The posterior probability of $H_0$ for the 24 data points of Table \ref{t4}}
\label{f2}
\end{figure}

As can be seen, for the hypotheses of Case $1$ and of Case $2$, based on the standard shown in Table \ref{t2}, our Bayesian model above has led to the potentially earlier 'termination' at the $17^{th}$-group data point with a total of $m=211$ 'events' of observed side effect of which $X_m=124$ were noted in the treatment group with an estimated relative risk of $\hat \gamma=1.4253$. The resulting Bayes Factors are $B_{211}^{(1)}=0.2901<0.3162$ and $B_{211}^{(2)}=0.3017<0.3162$. This leads to a rejection of $H_0^{(i)}$, $i=1, 2$, and the posterior probabilities of the null hypothesis and the alternative hypothesis given the data are
$$
\Pr(H_0^{(1)}: \gamma=1 \mid X_{211}=124)=0.2249,  \qquad \Pr(H_1^{(1)} :\gamma \neq 1 \mid X_{211}=124)=0.7751,
$$
and
$$
\Pr(H_0^{(2)} :\gamma= 1 \mid X_{211}=124)=0.2318, \qquad \Pr(H_1^{(2)} :\gamma > 1 \mid X_{211}=124)=0.7682,
$$
which are clearly smaller than the assumed prior probability $\pi_0=0.5$ of $H_0^{(i)}$, $i=1,2$, being true (marked in a red line in Figure \ref{f2}).

Furthermore, for the hypotheses of Case $3$, based on the standard shown in Table \ref{t2}, the $15^{th}$-group data point with a total of $m=190$ 'events' of observed side effect of which $X_m=107$ were noted in the treatment group, leading to estimated relative risk of $\hat \gamma=1.2892$. The resulting Bayes Factor is $B_{190}^{(3)}=0.1361<0.3162$ and the posterior probabilities of the null hypothesis and the alternative hypothesis given the data are 
$$
\Pr(H^{(3)}_0: \gamma \le 1 \mid X_{190}=107)=0.1198,  \qquad \Pr(H^{(3)}_1: \gamma > 1 \mid X_{190}=107)=0.8802,
$$
which is clearly smaller than the assumed prior probability $\pi_0=0.5$ of $H_0^{(3)}$ being true (marked in red line in Figure \ref{f2}).

\begin{table}[H]
\begin{center}{\footnotesize
\caption{The values of Bayes factor and posterior probability of $H_0$ for each data point among different cases with the informative prior.}
\label{t4}
\begin{tabular}{ c c c c c c c c c c }
\hline
Test & $m$ & $X_m$ & $\hat{\gamma}$ & $B_m^{(1)}$ & $\Pr(H_0^{(1)}\mid X_m)$ & $B_m^{(2)}$ & $\Pr(H_0^{(2)}\mid X_m)$ & $B_m^{(3)}$ & $\Pr(H_0^{(3)}\mid X_m)$ \\
\hline
1  & 12  & 1 &0.0909 &0.8321 &0.4542 &3.2157 &0.7628  &2.8646 &0.7412\\
2   &18   &5 &0.3846 &0.9115 &0.4768 &2.9928 &0.7495  &2.2834 &0.6954\\
3   &24  &11 &0.8462 &1.0431 &0.5106 &2.3194 &0.6987  &1.2234 &0.5502\\
4   &30  &15 &1.0000 &1.0640 &0.5155 &2.1280 &0.6803  &1.0000 &0.5000\\
5   &34  &15 &0.7895 &1.0398 &0.5098 &2.5854  &0.7211  &1.4864 &0.5978\\
6   &40  &17 &0.7391 &1.0137 &0.5034 &2.8419  &0.7397  &1.8036 &0.6433\\
7   &46  &20 &0.7692 &1.0265 &0.5065 &2.8658  &0.7413  &1.7917 &0.6418\\
8   &67  &34 &1.0303 &1.1360 &0.5318 &2.1710 &0.6846  &0.9111 &0.4767\\
9   &78 &39 &1.0000 &1.1590 &0.5368 &2.3181 &0.6986  &1.0000 &0.5000\\
10 &100  &44 &0.7857 &0.9627 &0.4905 &3.7991 &0.7916  &2.9465 &0.7466\\
11 &115  &51 &0.7969 &0.9583 &0.4894 &3.9771 &0.7991  &3.1501 &0.7590\\
12 &135  &63 &0.8750 &1.1289 &0.5303 &3.5494 &0.7802  &2.1442 &0.6820\\
13 &167  &88 &1.1139 &1.1885 &0.5431 &1.7609 &0.6378  &0.4817 &0.3251\\
14 &172  &91 &1.1235 &1.1693 &0.5390 &1.6904  &0.6283  &0.4456 &0.3083\\
15 &190 &107 &1.2892 &0.6786 &0.4043 &0.7710 &0.4354  &0.1361 &0.1198\\
16 &197 &113 &1.3452 &0.5061 &0.3360 &0.5497 &0.3547  &0.0862 &0.0794\\
17 &211 &124 &1.4253 &0.2901 &0.2249 &0.3017  &0.2318  &0.0400 &0.0384\\
18 &218 &130 &1.4773 &0.1920 &0.1611 &0.1966 &0.1643  &0.0237 &0.0231\\
19 &222 &134 &1.5227 &0.1325 &0.1170 &0.1345 &0.1185  &0.0151 &0.0149\\
20 &231 &141 &1.5667 &0.0824 &0.0761 &0.0831 &0.0767  &0.0086 &0.0085\\
21 &240 &148 &1.6087 &0.0494 &0.0471 &0.0496 &0.0473  &0.0048 &0.0047\\
22 &245 &153 &1.6630 &0.0276 &0.0269 &0.0277 &0.0269  &0.0025 &0.0025\\
23 &247 &155 &1.6848 &0.0216 &0.0212 &0.0217 &0.0212  &0.0019 &0.0019\\
24 &251 &157 &1.6702 &0.0225 &0.0220 &0.0225  &0.0220  &0.0020 &0.0019\\
\hline
\end{tabular}
}
\end{center}
\end{table} 

In conclusion, with the uniform prior, both Case $1$ and Case $2$ would reject and be 'terminated' at the $18^{th}$-group data point, and Case $3$ would reject and be 'terminated' at the $14^{th}$-group data point. However, with the informative prior we chose, both Case $1$ and Case $2$ would reject and be 'terminated' at the $17^{th}$-group data point, and Case $3$ would reject and be 'terminated' at the $15^{th}$-group data point.

\subsection{The Modified Bayesian Test} \label{s3.2}

\begin{figure}[H]
\centering
\begin{subfigure}[b]{0.328\textwidth}
\centering
\includegraphics[width=\textwidth]{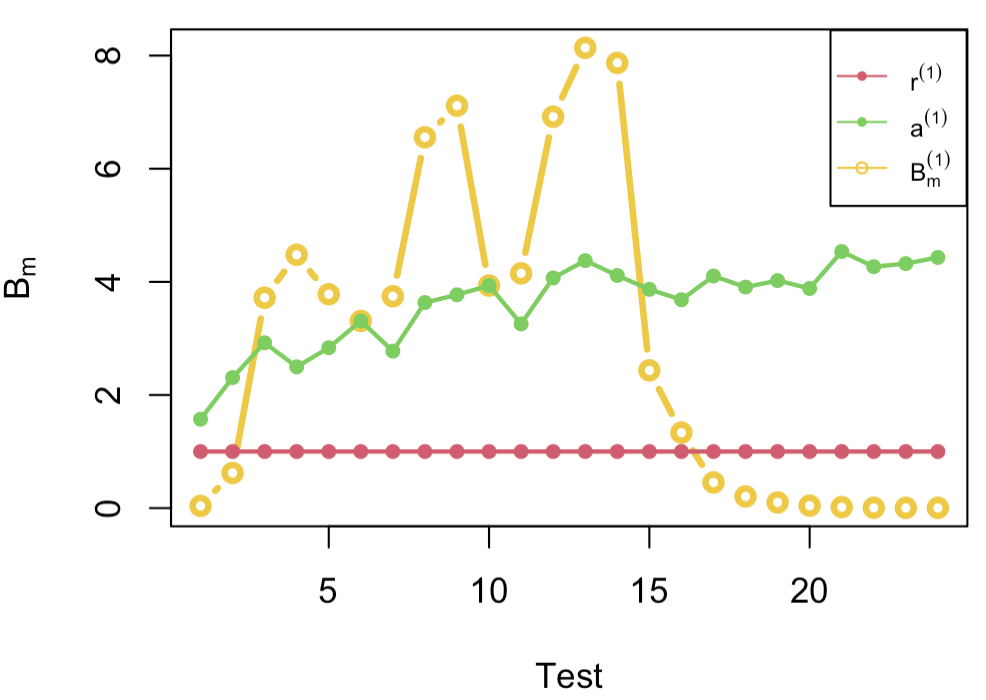}
\caption{Case 1}
\end{subfigure}
\hfill
\begin{subfigure}[b]{0.328\textwidth}
\centering
\includegraphics[width=\textwidth]{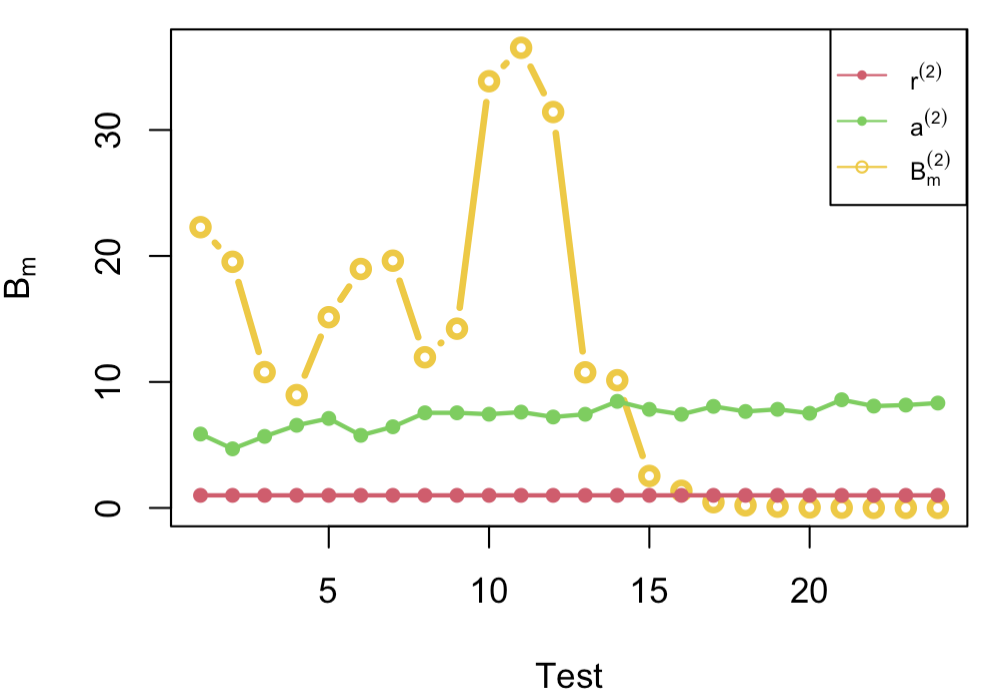}
\caption{Case 2}
\end{subfigure}
\begin{subfigure}[b]{0.328\textwidth}
\centering
\includegraphics[width=\textwidth]{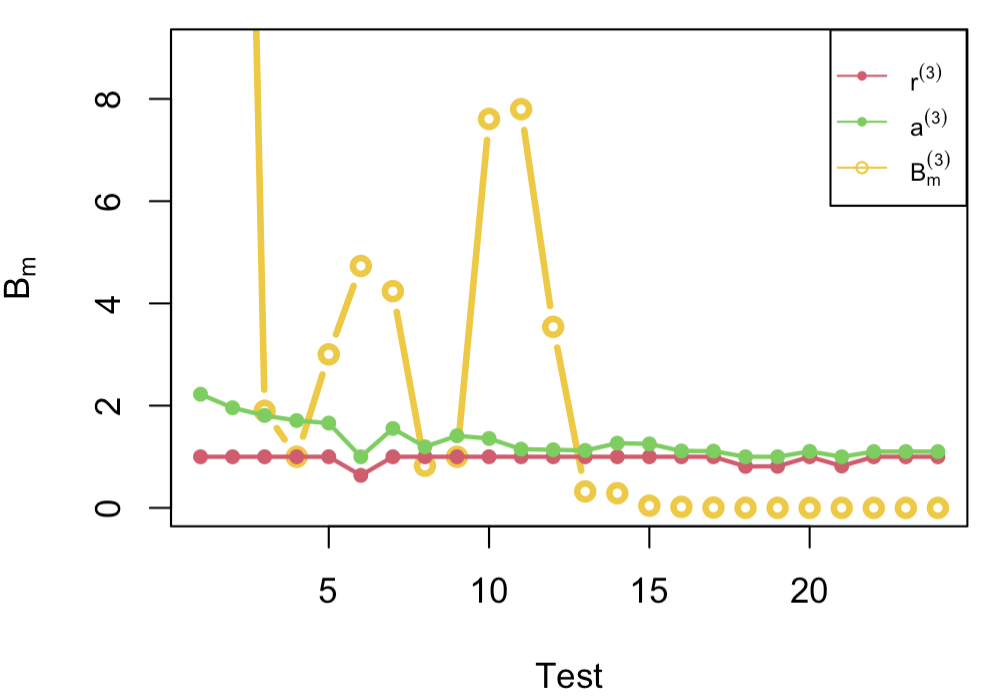}
\caption{Case 3}
\end{subfigure}
\caption{The acceptance and rejection boundaries of $B_m$ for the 24 data points of Table \ref{t5}}
\label{f3}
\end{figure}

\begin{figure}[H]
\centering
\begin{subfigure}[b]{0.328\textwidth}
\centering
\includegraphics[width=\textwidth]{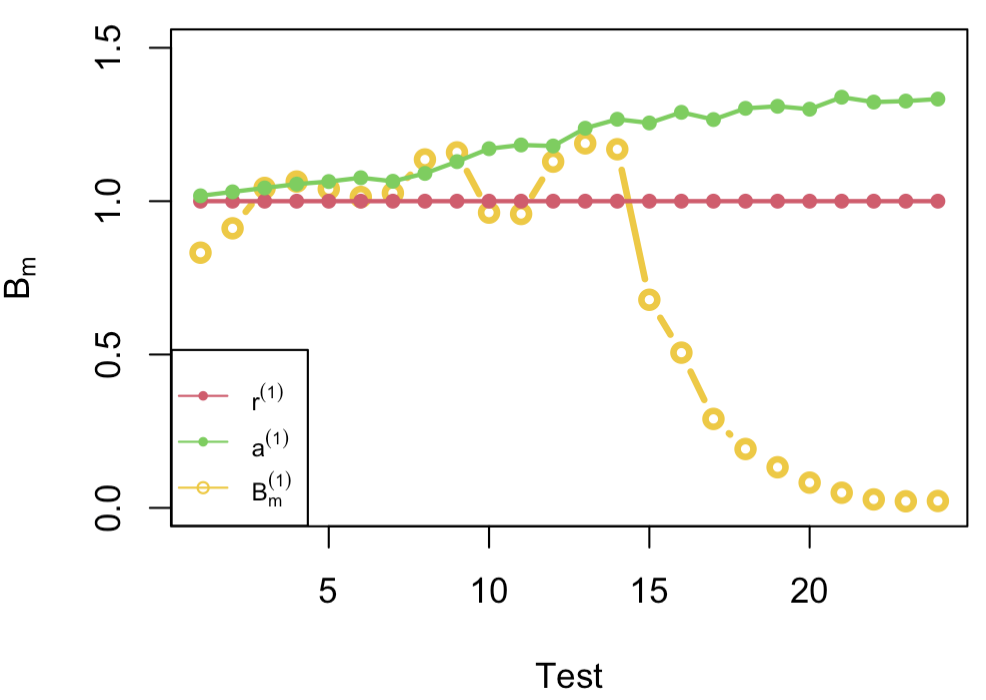}
\caption{Case 1}
\end{subfigure}
\hfill
\begin{subfigure}[b]{0.328\textwidth}
\centering
\includegraphics[width=\textwidth]{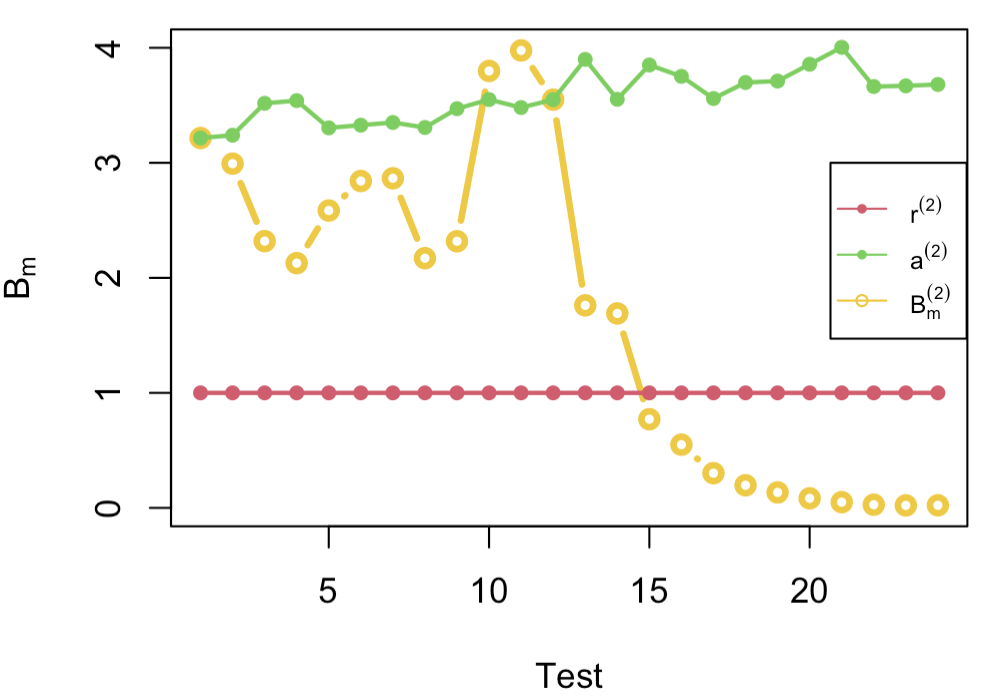}
\caption{Case 2}
\end{subfigure}
\begin{subfigure}[b]{0.328\textwidth}
\centering
\includegraphics[width=\textwidth]{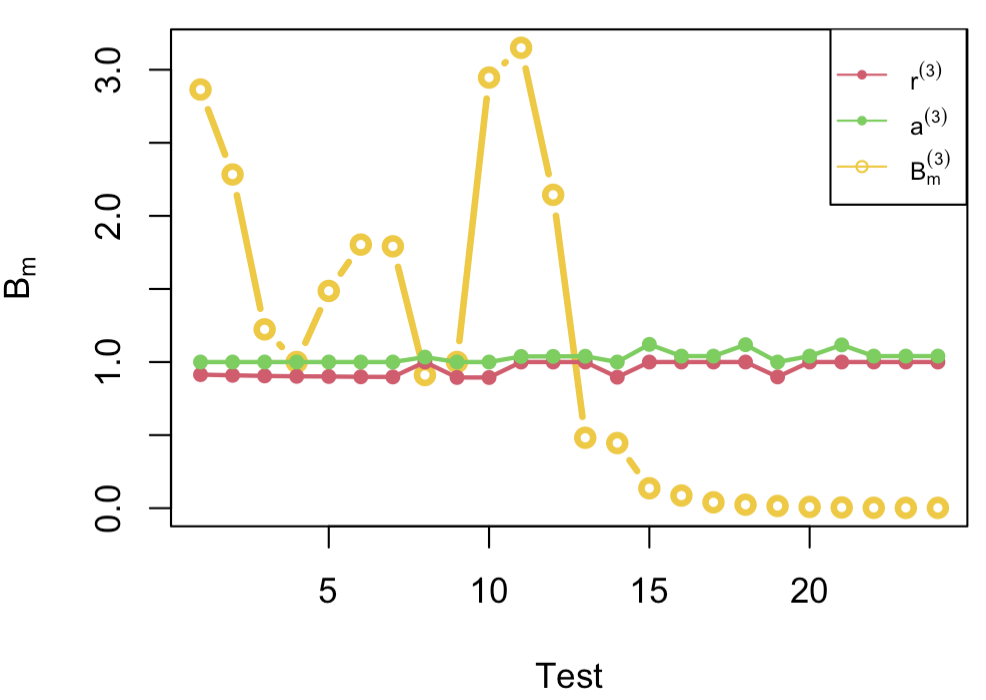}
\caption{Case 3}
\end{subfigure}
\caption{The acceptance and rejection boundaries of $B_m$ for the 24 data points of Table \ref{t7}}
\label{f4}
\end{figure}

Applying the modified Bayesian test in \eqref{2.21} to Case $1$, $2$, and $3$ along with the same prior structures we discussed above, we analyze the results of the above example again. Under both types of the priors, we obtain the same conclusions that were obtained utilizing the standard Bayesian test for the hypotheses $H_0^{(i)}$, $i=1,2,3$ (see Section \ref{s3.1}). The results corresponding to the non-informative (uniform) prior distributions are shown in Figure \ref{f3}, Table \ref{t5} and Table \ref{t6}. The results corresponding to the informative prior distributions are shown in Figure \ref{f4}, Table \ref{t7} and Table \ref{t8}. However, under both prior structures, utilizing the modified Bayesian test, we see that there are several cases with some data points falling into the 'no-decision' region. In all cases, we provide the calculated conditional Type~I and Type~II error probabilities, $\alpha^*$ and $\beta^*$, corresponding to the 'Reject' or 'Accept' decision (see Table \ref{t6} and Table \ref{t8}).

 One may interpret the results of Case $2$ under the informative prior case, for instance, as a guide for explaining the results shown in these $4$ tables. According to the results in Table \ref{t7}, at the $10^{th}$-group data point, since the Bayes Factor $B_{100}^{(2)}=3.7991$ is greater than $a^{(2)}=3.5500$, one would accept the null hypothesis that the relative risk is equal to $1$ and would report the corresponding conditional Type~II error probability $\beta^*=0.2136$ as shown in Table \ref{t8}. On the other hand, at the $17^{th}$-group data point, since $B_{211}^{(2)}=0.3017$ is less than $r^{(2)}=1$, along with the standard mentioned in Table \ref{t2}, we would reject the null hypothesis at the $17^{th}$-group data point and report the conditional Type~I error probability $\alpha^*=0.2318$. Additionally, there are some data points within 'no-decision' region since the Bayes Factor of the corresponding data point is within the values of $r^{(2)}$ and $a^{(2)}$.

%\medskip

\noindent I)  {\it{Non-informative (uniform) prior distribution}}

\begin{table}[H]
\begin{center}{\footnotesize
\caption{The values of Bayes factor and the boundaries based on it for each data point among different cases with the uniform prior.}
\label{t5}
\begin{tabular}{ c c c c c c c c c c c c c }
\hline
Test & $m$ & $X_m$ & $\hat{\gamma}$ & $B_m^{(1)}$ & $r^{(1)}$ &$a^{(1)}$ & $B_m^{(2)}$ &$r^{(2)}$ &$a^{(2)}$ & $B_m^{(3)}$ &$r^{(3)}$ &$a^{(3)}$ \\
\hline
1  & 12  & 1 &0.0909 &0.0381 &1 &1.5710 &22.2857  &1 &5.8652 &584.1429 &1 &2.2226 \\
2   &18   &5 &0.3846 &0.6210 &1 &2.3066  &19.5382 &1 &4.6903   &30.4623 &1 &1.9583 \\
3   &24  &11 &0.8462 &3.7195 &1 &2.9225  &10.7807 &1 &5.6789   &1.8984 &1 &1.8080 \\
4   &30  &15 &1.0000 &4.4784 &1 &2.4971   &8.9568 &1 &6.5607    &1.0000 &1 &1.7083 \\
5   &34  &15 &0.7895 &3.7811 &1 &2.8358  &15.1377 &1 &7.1038    &3.0035 &1 &1.6586 \\
6   &40  &17 &0.7391 &3.3088 &1 &3.3088  &18.9675 &1 &5.7628  &4.7325 &0.6363 &1 \\
7   &46  &20 &0.7692 &3.7458 &1 &2.7747  &19.6273 &1 &6.4415  &4.2398 &1 &1.5532 \\
8   &67  &34 &1.0303 &6.5554 &1 &3.6340  &11.9580 &1 &7.5499   &0.8241 &1 &1.1945 \\
9   &78 &39 &1.0000 &7.1142 &1 &3.7717  &14.2285  &1 &7.5508  &1.0000 &1 &1.4105 \\
10 &100  &44 &0.7857 &3.9342 &1 &3.9342 &33.8717  &1 &7.4389   &7.6095 &1 &1.3574 \\
11 &115  &51 &0.7969 &4.1504 &1 &3.2564  &36.5229 &1 &7.6064    &7.7999 &1 &1.1485 \\
12 &135  &63 &0.8750 &6.9197 &1 &4.0694  &31.4257 &1 &7.2192   &3.5415 &1 &1.1371 \\
13 &167  &88 &1.1139 &8.1376 &1 &4.3758  &10.7609 &1 &7.4343  &0.3224 &1 &1.1233 \\
14 &172  &91 &1.1235 &7.8705 &1 &4.1132  &10.1371 &1 &8.4590    &0.2880 &1 &1.2656 \\
15 &190 &107 &1.2892 &2.4347 &1 &3.8676   &2.5390 &1 &7.8237   &0.0429 &1 &1.2517 \\
16 &197 &113 &1.3452 &1.3341 &1 &3.6836   &1.3606 &1 &7.4293 &0.0199 &1 &1.1136 \\
17 &211 &124 &1.4253 &0.4531 &1 &4.1037   &0.4556 &1 &8.0592  &0.0055 &1 &1.1098 \\
18 &218 &130 &1.4773 &0.2055 &1 &3.9079  &0.2059  &1 &7.6559  &0.0022 &0.8116 &1 \\
19 &222 &134 &1.5227 &0.1003 &1 &4.0227 &0.1004   &1 &7.8268    &0.0010 &0.8131 &1 \\
20 &231 &141 &1.5667 &0.0427 &1 &3.8834  &0.0427  &1 &7.5158    &0.0004 &1 &1.1049 \\
21 &240 &148 &1.6087 &0.0175 &1 &4.5363  &0.0175  &1 &8.5869   &0.0001 &0.8193 &1 \\
22 &245 &153 &1.6630 &0.0060 &1 &4.2679   &0.0060 &1 &8.0842    &0.0000 &1 &1.1018 \\
23 &247 &155 &1.6848 &0.0039 &1 &4.3227 &0.0039   &1 &8.1648    &0.0000 &1 &1.1014 \\
24 &251 &157 &1.6702 &0.0044 &1 &4.4320  &0.0044  &1 &8.3256   &0.0000 &1 &1.1006 \\
\hline
\end{tabular}
}
\end{center}
\end{table}

\begin{table}[H]
\begin{center}{\scriptsize 
\caption{The decision and corresponding conditional $\alpha^*$ and $\beta^*$ for each data point among different cases with the uniform prior. (Decision: 'R' indicates Reject, 'A' indicates Accept, 'ND' indicates No Decision, 'NA' indicates No Answer.)}
\label{t6}
\begin{tabular}{ c c c c c c c c c c c }
\hline
Test & $\hat{\gamma}$ & $B_m^{(1)}$ & Decision &$\alpha^*(\text{or } \beta^*)$ & $B_m^{(2)}$ & Decision &$\alpha^*(\text{or } \beta^*)$ & $B_m^{(3)}$ & Decision &$\alpha^*(\text{or } \beta^*)$ \\
\hline
1   &0.0909 &0.0381 &R &0.0367 &22.2857 &A &0.0441 &584.1429 &A &0.0016 \\
2    &0.3846 &0.6210 &R &0.3831  &19.5382 &A &0.0491   &30.4623 &A &0.0261 \\
3    &0.8462 &3.7195 &A &0.2145  &10.7807 &A &0.0926   &1.8984 &A &0.3313 \\
4    &1.0000 &4.4784 &A &0.1841   &8.9568 &A &0.1072   &1.0000 &ND &NA \\
5    &0.7895 &3.7811 &A &0.2145  &15.1377 &A &0.0621    &3.0035 &A &0.2460\\
6    &0.7391 &3.3088 &ND &NA  &18.9675 &A &0.0551  &4.7325 &A &0.1742 \\
7   &0.7692 &3.7458 &A &0.2145 &19.6273 &A &0.0491 &4.2398 &A &0.1742 \\
8    &1.0303 &6.5554 &A &0.1391 &11.9580 &A &0.0805   &0.8241 &R &0.4518 \\
9    &1.0000 &7.1142 &A &0.1391  &14.2285  &A &0.0705  &1.0000 &ND &NA \\
10  &0.7857 &3.9342 &ND &NA &33.8717  &A &0.0303   &7.6095 &A &0.1173 \\
11  &0.7969 &4.1504 &A &0.2145  &36.5229 &A &0.0279  &7.7999 &A &0.0944 \\
12  &0.8750 &6.9197 &A &0.1391  &31.4257 &A &0.0330  &3.5415 &A &0.2082 \\
13  &1.1139 &8.1376 &A &0.1098  &10.7609 &A &0.0926  &0.3224 &R &0.2438 \\
14  &1.1235 &7.8705 &A &0.1228  &10.1371 &A &0.0926    &0.2880 &R &0.2236 \\
15  &1.2892 &2.4347 &ND &NA  &2.5390 &ND &NA  &0.0429 &R &0.0411 \\
16  &1.3452 &1.3341 &ND &NA   &1.3606 &ND &NA &0.0199 &R &0.0195 \\
17  &1.4253 &0.4531 &R &0.3118   &0.4556 &R &0.3130  &0.0055 &R &0.0055 \\
18  &1.4773 &0.2055 &R &0.1705  &0.2059  &R &0.1707  &0.0022 &R &0.0022\\
19  &1.5227 &0.1003 &R &0.0912 &0.1004  &R &0.0912 &0.0010 &R &0.0010 \\
20  &1.5667 &0.0427 &R &0.0410  &0.0427 &R &0.0410    &0.0004 &R &0.0004 \\
21  &1.6087 &0.0175 &R &0.0172  &0.0175  &R &0.0172   &0.0001 &R &0.0001\\
22  &1.6630 &0.0060 &R &0.0060  &0.0060 &R &0.0060   &0.0000 &R &0.0000\\
23  &1.6848 &0.0039 &R &0.0039 &0.0039   &R &0.0039    &0.0000 &R &0.0000\\
24  &1.6702 &0.0044 &R &0.0044  &0.0044  &R &0.0044 &0.0000 &R &0.0000 \\
\hline
\end{tabular}
}
\end{center}
\end{table}

\medskip
 
\noindent II)  {\it{Informative prior distribution}}

\begin{table}[H]
\begin{center}{\footnotesize
\caption{The values of Bayes factor and the decision based on it for each data point among different cases with the informative prior.}
\label{t7}
\begin{tabular}{ c c c c c c c c c c c c c }
\hline
Test & $m$ & $X_m$ & $\hat{\gamma}$ & $B_m^{(1)}$ & $r^{(1)}$ &$a^{(1)}$ & $B_m^{(2)}$ & $r^{(2)}$ &$a^{(2)}$ & $B_m^{(3)}$ & $r^{(3)}$ &$a^{(3)}$ \\
\hline
1  & 12  & 1 &0.0909 &0.8321 &1 &1.0175 &3.2157 &1 &3.2157  &2.8646 &0.9138 &1 \\
2   &18   &5 &0.3846 &0.9115 &1 &1.0304 &2.9928 &1 &3.2401 &2.2834 &0.9091 &1\\
3   &24  &11 &0.8462 &1.0431 &1 &1.0431 &2.3194 &1 &3.5175  &1.2234 &0.9050 &1 \\
4   &30  &15 &1.0000 &1.0640 &1 &1.0557 &2.1280 &1 &3.5410  &1.0000 &0.9020 &1 \\
5   &34  &15 &0.7895 &1.0398 &1 &1.0640 &2.5854  &1 &3.3040  &1.4864 &0.9010 &1 \\
6   &40  &17 &0.7391 &1.0137 &1 &1.0764 &2.8419  &1 &3.3275  &1.8036 &0.8988 &1 \\
7   &46  &20 &0.7692 &1.0265 &1 &1.0649 &2.8658  &1 &3.3507  &1.7917 &0.8980 &1 \\
8   &67  &34 &1.0303 &1.1360 &1 &1.0905 &2.1710 &1 &3.3071  &0.9111 &1 &1.0337 \\
9   &78 &39 &1.0000 &1.1590 &1 &1.1290 &2.3181 &1  &3.4708  &1.0000 &0.8943 &1 \\
10 &100  &44 &0.7857 &0.9627 &1 &1.1711 &3.7991 &1 &3.5500  &2.9465 &0.8940 &1 \\
11 &115  &51 &0.7969 &0.9583 &1 &1.1832 &3.9771 &1 &3.4806  &3.1501 &1 &1.0379 \\
12 &135  &63 &0.8750 &1.1289 &1 &1.1799 &3.5494 &1 &3.5494  &2.1442 &1 &1.0384 \\
13 &167  &88 &1.1139 &1.1885 &1 &1.2378 &1.7609 &1 &3.8995  &0.4817 &1 &1.0395 \\
14 &172  &91 &1.1235 &1.1693 &1 &1.2671 &1.6904  &1 &3.5525  &0.4456 &0.8962 &1 \\
15 &190 &107 &1.2892 &0.6786 &1 &1.2550 &0.7710 &1 &3.8506  &0.1361 &1 &1.1207 \\
16 &197 &113 &1.3452 &0.5061 &1 &1.2897 &0.5497 &1 &3.7520  &0.0862 &1 &1.0403 \\
17 &211 &124 &1.4253 &0.2901 &1 &1.2661 &0.3017  &1 &3.5590  &0.0400 &1 &1.0404 \\
18 &218 &130 &1.4773 &0.1920 &1 &1.3027 &0.1966 &1 &3.6984 &0.0237 &1 &1.1186 \\
19 &222 &134 &1.5227 &0.1325 &1 &1.3094 &0.1345  &1 &3.7108  &0.0151 &0.8981 &1 \\
20 &231 &141 &1.5667 &0.0824 &1 &1.2999 &0.0831  &1 &3.8571 &0.0086 &1 &1.0403 \\
21 &240 &148 &1.6087 &0.0494 &1 &1.3390 &0.0496 &1 &4.0043  &0.0048 &1 &1.1172 \\
22 &245 &153 &1.6630 &0.0276 &1 &1.3231 &0.0277 &1 &3.6634 &0.0025 &1 &1.0404 \\
23 &247 &155 &1.6848 &0.0216 &1 &1.3264 &0.0217 &1 &3.6694  &0.0019 &1 &1.0404 \\
24 &251 &157 &1.6702 &0.0225 &1 &1.3329 &0.0225  &1 &3.6814 &0.0020 &1 &1.0406 \\
\hline
\end{tabular}
}
\end{center}
\end{table}

\begin{table}[H]
\begin{center}{\scriptsize 
\caption{The decision and corresponding conditional $\alpha^*$ and $\beta^*$ for each data point among different cases with the informative prior. (Decision: 'R' indicates Reject, 'A' indicates Accept, 'ND' indicates No Decision, 'NA' indicates No Answer.)}
\label{t8}
\begin{tabular}{ c c c c c c c c c c c }
\hline
Test & $\hat{\gamma}$ & $B_m^{(1)}$ & Decision &$\alpha^*(\text{or } \beta^*)$ & $B_m^{(2)}$ & Decision &$\alpha^*(\text{or } \beta^*)$ & $B_m^{(3)}$ & Decision &$\alpha^*(\text{or } \beta^*)$ \\
\hline
1   &0.0909 &0.8321 &R &0.4542 &3.2157 &ND &NA   &2.8646 &A &0.2610  \\
2    &0.3846 &0.9115 &R &0.4769  &2.9928 &ND &NA &2.2834 &A &0.2903 \\
3   &0.8462 &1.0431 &ND &NA  &2.3194 &ND &NA   &1.2234 &A &0.4549  \\
4   &1.0000 &1.0640 &A &0.4287  &2.1280 &ND &NA   &1.0000 &ND &NA  \\
5    &0.7895 &1.0398 &ND &NA  &2.5854  &ND &NA  &1.4864 &A &0.3862  \\
6   &0.7391 &1.0137 &ND &NA  &2.8419  &ND &NA   &1.8036 &A &0.3530  \\
7    &0.7692 &1.0265 &ND &NA  &2.8658  &ND &NA  &1.7917 &A &0.3530  \\
8    &1.0303 &1.1360 &A &0.4389  &2.1710 &ND &NA  &0.9111 &R & 0.4767 \\
9    &1.0000 &1.1590 &A &0.4389  &2.3181 &ND &NA   &1.0000 &ND &NA  \\
10  &0.7857 &0.9627 &R &0.4905  &3.7991 &A &0.2136  &2.9465 &A &0.2610  \\
11  &0.7969 &0.9583 &R &0.4894  &3.9771 &A &0.2034  &3.1501 &A &0.2334  \\
12  &0.8750 &1.1289 &ND &NA  &3.5494 &ND &NA   &2.1442 &A &0.3210  \\
13  &1.1139 &1.1885 &ND &NA  &1.7609 &ND &NA &0.4817 &R &0.3251 \\
14  &1.1235 &1.1693 &ND &NA  &1.6904 &ND &NA  &0.4456 &R &0.3082 \\
15  &1.2892 &0.6786 &R &0.4043  &0.7710 &R &0.4353  &0.1361 &R &0.1198 \\
16  &1.3452 &0.5061 &R &0.3360  &0.5497 &R &0.3547  &0.0862 &R &0.0794 \\
17  &1.4253 &0.2901 &R &0.2249  &0.3017 &R &0.2318   &0.0400 &R &0.0385  \\
18  &1.4773 &0.1920 &R &0.1611  &0.1966 &R &0.1643 &0.0237 &R & 0.0232 \\
19  &1.5227 &0.1325 &R &0.1170  &0.1345  &R &0.1186   &0.0151 &R &0.0149  \\
20  &1.5667 &0.0824 &R &0.0761  &0.0831  &R &0.0767  &0.0086 &R &0.0085 \\
21  &1.6087 &0.0494 &R &0.0471  &0.0496 &R &0.0473   &0.0048 &R &0.0048  \\
22  &1.6630 &0.0276 &R &0.0269  &0.0277 &R &0.0270 &0.0025 &R &0.0025 \\
23  &1.6848 &0.0216 &R &0.0211  &0.0217 &R &0.0212  &0.0019 &R &0.0019 \\
24  &1.6702 &0.0225 &R &0.0220  &0.0225  &R &0.0220  &0.0020 &R &0.0020  \\
\hline
\end{tabular}
}
\end{center}
\end{table} 

\subsection{HPD Region}\label{s3.3}

Once the 'termination' point of the study is known (with a given prior distribution), it is possible to obtain the $(1-c)\times 100 \%$ Highest Posterior Density (HPD) region for the RR, $\gamma$, as now one has available its posterior $pdf$ given the data. We illustrate this by calculating a $95 \% $ HPD region for each 'terminated' data point we discuss above.

$\bullet$ {\bf Case 1 ($H_0^{(1)}: \gamma = 1 \text{ against } H_1^{(1)}:\gamma \neq 1$) $\&$ Case 2 ($H_0^{(2)}: \gamma = 1 \text{ against } H_1^{(2)}:\gamma > 1$):}

\begin{figure}[H]
\centering
\includegraphics[scale=0.4]{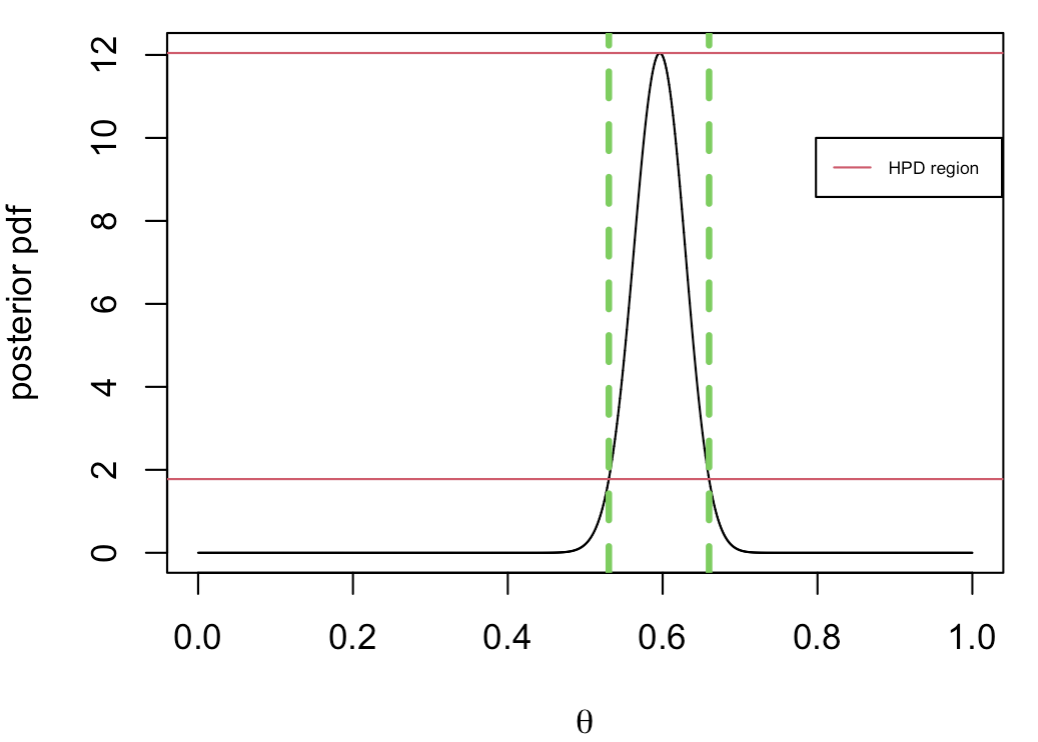}
\caption{The $95 \%$ HPD region of $\theta\mid X_{218} \sim \mathcal{B}eta (131,89)$ where $X_{218}=130$ at $18^{th}$-group data point with the posterior $pdf$ of $\theta\mid X_{218}$.}
\label{f5}
\end{figure}

At the $18^{th}$-group data point, which is the 'terminated' data point for Case $1$ and Case $2$ with the uniform prior, we obtain the posterior distribution of $\theta \mid X_{218} \sim \mathcal{B}eta (131,89)$ where $X_{218}=130$. The calculated $95 \% $ HPD region of $\theta$ is $(0.5305,0.6599)$ shown in Figure \ref{f5}, which indicates the corresponding $95 \% $ HPD region of $\gamma$, the Relative Risk, is $(1.1298,1.9407)$. Notice that the conditional Type~I error probability is $0.1705$ for Case $1$ and is $0.1707$ for Case $2$ under the modified Bayesian test. Although the $95 \% $ HPD region excludes $\gamma=1$, the conditional Type~I error probabilities indicate there are around $17 \%$ in favor of the null hypothesis for Case $1$ and $2$.

At the $17^{th}$-group data point, which is the 'terminated' data point for Case $1$ and Case $2$ with the informative prior, we obtain the posterior distribution of $\theta \mid X_{211} \sim \mathcal{B}eta (237.8288,200.8288)$ where $X_{211}=124$. The calculated $95 \% $ HPD region of $\theta$ is $(0.4955,0.5888)$ which indicates the corresponding $95 \% $ HPD region of $\gamma$ is $(0.9820,1.4318)$. Notice that the conditional Type~I error probability is $0.2249$ for Case $1$ and is $0.2318$ for Case $2$ under the modified Bayesian test.

$\bullet$ {\bf Case 3 ($H_0^{(3)}: \gamma \le 1 \text{ against } H_1^{(3)}:\gamma > 1$):}

At the $14^{th}$-group data point, which is the 'terminated' data point for Case $3$ with the uniform prior, we obtain the posterior distribution of $\theta \mid X_{172} \sim \mathcal{B}eta (92,82)$ where $X_{172}=91$. The calculated $95 \% $ HPD region of $\theta$ is $(0.4546,0.6027)$ which indicates the corresponding $95 \% $ HPD region of $\gamma$ is $(0.8336,1.5168)$. The conditional Type~I error probability is $0.2236$ under the modified Bayesian test.

At the $15^{th}$-group data point, which is the 'terminated' data point for Case $3$ with the informative prior, we obtain the posterior distribution of $\theta \mid X_{190} \sim \mathcal{B}eta (220.8288,196.8288)$ where $X_{190}=107$. The calculated $95 \% $ HPD region of $\theta$ is $(0.4809,0.5765)$ which indicates the corresponding $95 \% $ HPD region of $\gamma$ is $(0.9263,1.3613)$. The conditional Type~I error probability is $0.1198$ under the modified Bayesian test.

\begin{rem} \label{r3}
We notice that some $95\%$ HPD regions may include $\gamma=1$, the value of $\gamma$ under the null hypothesis. Since the Bayes Factor shows that the overall evidence from the data favors the alternative compared to the null, the $95\%$ HPD region reflects that the null is not entirely implausible. Even if the null value is within the $95\%$ HPD region, the posterior may still assign higher credibility to other values (those outside the null), which explains why the Bayes Factor rejects the null. Hence, the null hypothesis may still be within a credible interval, but its likelihood is overshadowed by the alternative. It also indicates that the null hypothesis is rejected substantially.

\end{rem}

\subsection{Connection to the UMPBT}\label{s3.4}

\citet{johnson2013uniformly} has introduced the new concept of the Uniformly Most Powerful Bayesian Test (UMPBT) and proved its existence in \citet{nikooienejad2021existence}. Considering Case $2$, now treated as the UMPBT, we construct the corresponding 'group' sequential Bayesian test with a certain 'termination' point. However, we note that in the current context, the threshold of the Bayes Factor $B_m$ to reject the null hypothesis is of the form $B_m < 1/\lambda$ for some $\lambda>0$. By Lemma $1$ in \citet{johnson2013uniformly}, the UMPBT($1/\lambda$) is constructed by finding $\theta_1$ that $\theta_1>\theta_0$ and
$$
\theta_1=\arg \min_{\theta} \frac{\ln{(1/\lambda)}-m \left[ \ln{(1-\theta)}-\ln{(1-\theta_0)} \right]}{\ln{[\theta/(1-\theta)]}-\ln{[\theta_0/(1-\theta_0)]}}, 
$$
which can be solved numerically. For instance, by applying the suggestion of the Jeffrey's evidence level shown in Table \ref{t2}, we may reject the null hypothesis if $B_m<0.3162$, so that $1/\lambda = 0.3162$ in this case.

\noindent I)  {\it{Non-informative (uniform) prior distribution}}

Since the 'termination' point of Case $2$ with the uniform prior is the $18^{th}$-group data point, we have $B_{218}=0.2059$ and $\Pr(H_0^{(2)} \mid X_{218}=130)=0.1707$, $m=218$ and $X_m=130$. We use $1/\lambda=0.2059$ and obtain the corresponding $\theta_1=0.5601$. At this value of $\theta_1$ and $\theta_0=0.5$, $B_m<1/\lambda$ whenever $123$ or more of the $n=218$ patients enrolled in this fixed sample size study. Accordingly, for the fixed sample size test (with $n=218$), once the number of the observed side effect $X_n \ge 123$, we would reject the null hypothesis, incurring a (classical) Type~I error probability $\alpha=\operatorname{P}_{\theta_0}(X_n \ge 123)=0.0336$. This rejection region for this $3.36\%$ significance test also corresponds to the region for which the Bayes Factor corresponding to the UMPBT($1/\lambda$) for all values of $1/\lambda \in (0.166, 0.2114)$. With equal prior probabilities of the null and the alternative hypotheses we assumed previously ($\ell=1$), the implied significant level of $0.0336$ for this UMPBT which approximately corresponds to the Bayesian test resulting with
$$
0.1424 < \Pr(H_0^{(2)} \mid Data)<0.1745.
$$
This serves as illustration to the fact that unlike the posterior probabilities, the p-value or attained significant level, is not a true measure of the strength of the evidence in favor/against the null hypothesis, see \citet{sellke2001calibration} for a discussion of this point.

However, note that, if we choose Jeffrey's evidence level of $1/ \lambda=0.3162$, the calculated $B_{218}=0.2059$ and $\Pr(H_0^{(2)} \mid X_{218}=130)=0.1707$ which would not match the results obtained from the UMPBT procedure in this case. This may be attributed to the group sequential nature of the data and the uniform prior that we considered here.

\noindent II)  {\it{Informative prior distribution}}

Since the 'termination' point of Case $2$ with the informative prior is the $17^{th}$-group data point, we have $B_{211}=0.3017$ and $\Pr(H_0^{(2)} \mid X_{211}=124)=0.2318$, $m=211$ and $X_m=124$. Applying Jeffrey's evidence level of $1/\lambda=0.3162$, with the corresponding (equivalent) fixed sample size of this UMPBT $n=211$, we obtain the corresponding $\theta_1=0.5522$. With this value of $\theta_1$ and $\theta_0=0.5$, $B_m < 1/\lambda$ whenever $117$ or more of the $n=211$ patients enrolled in this fixed sample size study. Accordingly, for the fixed sample size test, once the number of the observed side effect $X_n \ge 117$, we would reject the null hypothesis, incurring a (classical) Type~I error $\alpha=\operatorname{P}_{\theta_0}(X_n \ge 117)=0.0648$. This rejection region for this $6.48\%$ significance test also corresponds to the region for which the Bayes Factor corresponding to the UMPBT($1/\lambda$) for all values of $1/\lambda \in (0.2854, 0.3520)$. With equal prior probabilities of the null and the alternative hypotheses we assumed previously ($\ell=1$), the implied significant level of $0.0648$ for this UMPBT which approximately corresponds to the Bayesian test resulting with
$$
0.2221 <\Pr(H_0^{(2)} \mid Data)<  0.2603.
$$

\section{Summary and Discussion}

In this paper, we propose the standard Bayesian test and the modified Bayesian test of hypotheses concerning the Relative Risk, $\gamma$, of a two-arm clinical trial. In this context, the Relative Risk $\gamma$ can be represented as $\theta(\gamma)$, a parameter representing an event probability. Since the output of each observation collected is binary data, the sequential experimental process can be viewed as a sequential binomial process with probability $\theta(\gamma)$. Note that, under the Bayesian framework, based on the SRP, each data point remains unaffected by the optimal stopping rule used in the process (unlike the 'classical' sequential test). Within the Bayesian framework, we consider the sequential testing of hypotheses concerning the parameter $\theta(\gamma)$. By utilizing the conjugate beta-binomial model, we are able to obtain the corresponding decision 'rule' for the test at each observed data point as determined from the calculated value of the Bayes Factor.

To deal with the values of the Bayes Factor that are too close to the decision boundaries, Jefferey's criteria for evidentiary level for a 'rejection of $H_0$' are taken into consideration. Additionally, we consider a modification of the standard Bayesian test (\citet{berger1997unified}) which includes a 'no-decision' region and calculate corresponding conditional Type~I and Type~II error probabilities $\alpha^*$ and $\beta^*$. To illustrate our approach and the methods discussed, we analyze the data presented in \citet{silva2020optimal} under the three different hypothesis testing cases, and under several different priors. It is worth pointing out that the informative prior we proposed is designed to use the knowledge of the bounded-width probability. For each case with a different priors, we analyze the calculated Bayes Factor corresponding to each data point and make the decision within two Bayesian tests.

Specifically, we consider in Case $2$ the very same hypothesis testing problem as shown in \citet{silva2020optimal}. The result presented in \citet{silva2020optimal} indicated that the testing procedure should have been terminated at the $19^{th}$-group data point. However, the results obtained from both the standard Bayesian test and the modified Bayesian test lead to a potentially earlier 'termination' at the $18^{th}$-group data point under the uniform prior and at the $17^{th}$-group data point under an informative prior. Furthermore, utilizing the modified Bayesian test with a uniform prior, at the $18^{th}$-group data point, the Type~I error conditional probability is $\alpha^*=0.1707$ (given the data). Similarly, with the modified Bayesian test and under an informative prior, at the $17^{th}$-group data point, the conditional Type~I error probability is $\alpha^*=0.2318$ (given the data). From Section \ref{s3.4}, we find our 'termination' based on the Bayesian test with a uniform prior corresponding to a p-value of $0.0336$ and with an informative prior corresponding to a p-value of $0.0648$ from the frequentist point of view.

Case $3$ has the very same hypothesis testing problem as shown in \citet{silva2022bounded}. The result presented in \citet{silva2022bounded} indicated that the testing procedure should have been terminated at the $18^{th}$-group data point. However, the results obtained from both the standard Bayesian test and the modified Bayesian test lead to the potentially earlier 'termination' at the $14^{th}$-group data point with a uniform prior and at the $15^{th}$-group data point with an informative prior. Furthermore, with the modified Bayesian test under a uniform prior, at the $14^{th}$-group data point, the conditional Type~I error probability $\alpha^*=0.2236$ (given the data). Similarly, with the modified Bayesian test under an informative prior, at the $15^{th}$-group data point, the conditional Type~I error probability $\alpha^*=0.1198$ (given the data).

These cases illustrate the benefits of Bayesian testing in this context. Unlike the classical sequential test, which typically requires heavy calculations to determine 'stopping boundaries' as well as 'decision boundaries' (for a-priory specified Type~I and Type~II error probabilities), the Bayesian test is easy to use and implement, and it could obtain the test results in a relatively simple manner. Compared with the results obtained in \citet{silva2020optimal} and \citet{silva2022bounded}, we would conclude that our results are consistent with their results. Although our decisions for Case $2$ and Case $3$ arrive slightly earlier than theirs, the values of the data points are similar. In addition, we are able to illustrate, using the very same data, the connection of our Bayesian test and the UMPBT. With this connection, the results we obtained from the Bayesian test can be approximately interpreted by the rejection region and the p-value from the frequentist perspective of the UMP. Moreover, by applying modified Bayesian test, we could even obtain conditional error probabilities as well, which give us a measure of the data evidence in making decisions (of 'Reject' $H_0$ or 'Accept' $H_0$). Indeed, for such binary data hypothesis testing, utilizing Bayesian framework could provide a straightforward and easy path to analyze the data.

\printbibliography

\section{Appendix}

In this section, we present the results obtained from the standard Bayesian test and the modified Bayesian test under the Jeffrey's (non-informative) prior of the data from \citet{silva2020optimal}.

\subsection{The Standard Bayesian Test}\label{s5.1}

Introduced in \citet{jeffreys1946invariant}, the Jeffrey's prior as considered an 'objective' or non-informative prior distribution is $\mathcal{B}eta \sim (1/2,1/2)$ for binomial distribution in our case. We calculated the Bayes factor $B^{(i)}_m$ and the corresponding posterior probability $Pr(H_0^{(i)} \mid X_m=x)$, $i=1, 2, 3$, for the hypotheses tests on the RR, $\gamma$, in the three cases we considered. The numerical results are provided in Table \ref{tj1} below.

\begin{figure}[H]
\centering
\begin{subfigure}[b]{0.328\textwidth}
\centering
\includegraphics[width=\textwidth]{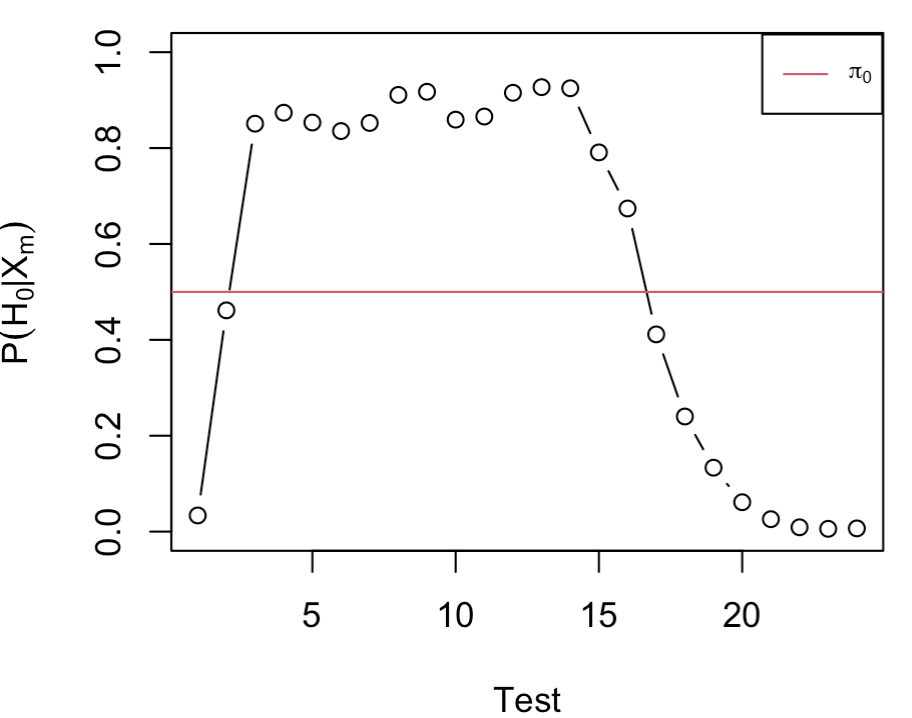}
\caption{Case 1}
\end{subfigure}
\hfill
\begin{subfigure}[b]{0.328\textwidth}
\centering
\includegraphics[width=\textwidth]{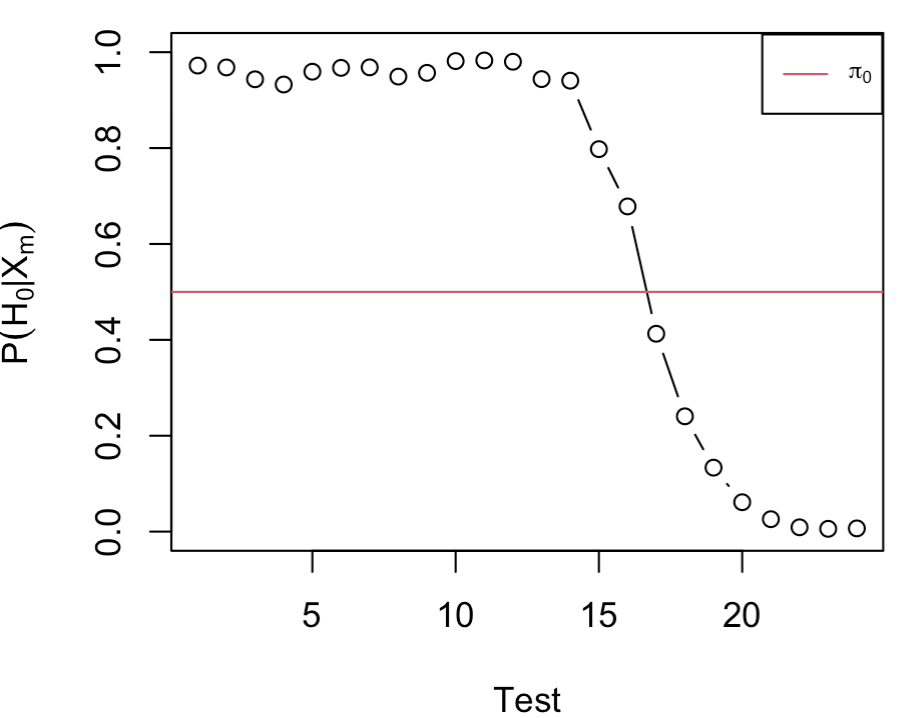}
\caption{Case 2}
\end{subfigure}
\begin{subfigure}[b]{0.328\textwidth}
\centering
\includegraphics[width=\textwidth]{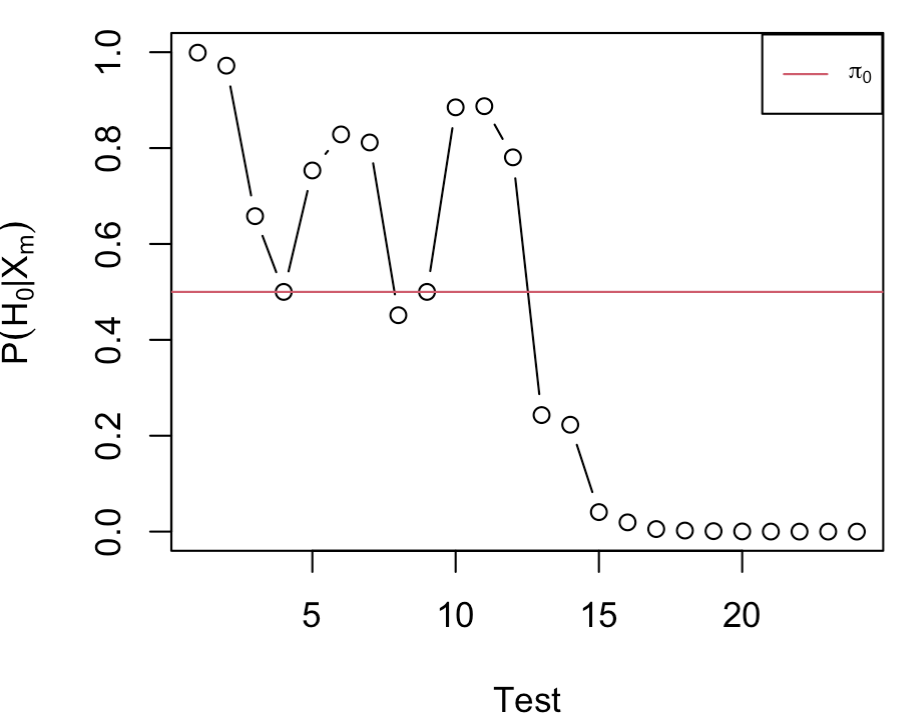}
\caption{Case 3}
\end{subfigure}
\caption{The posterior probability of $H_0$ for the 24 data points of Table \ref{tj1}}
\label{fj1}
\end{figure}

As can be seen, for the hypotheses of Case $1$, our Bayesian model has led to the potentially earlier 'termination' at the $18^{th}$-group data point with a total of $m=218$ 'events' of observed side effect of which $X_m=130$ were noted in the treatment group with an estimated relative risk of $\hat \gamma=X_m/(m-X_m)=1.4773$. The resulting Bayes Factor is $B_{218}^{(1)}=0.3160$, which is less than $0.3162$ based on the standard shown in Table \ref{t2}. This leads to a rejection of $H_0^{(1)}$ and the posterior probabilities of the null hypothesis and the alternative hypothesis given the data are
$$
\Pr(H_0^{(1)}: \gamma=1 \mid X_{218}=130)=0.2401  \qquad \Pr(H_1^{(1)} :\gamma \neq 1 \mid X_{218}=130)=0.7599,
$$
which are clearly smaller than the assumed prior probability $\pi_0=0.5$ of $H_0^{(1)}$ being true (marked in a red line in Figure \ref{fj1}).

For the hypotheses of Case $2$, one would 'terminate' the observation process at the $19^{th}$-group data point with a total of $m=222$ 'events' of observed side effect of which $X_m=134$ were noted in the treatment group, leading to the estimated relative risk of $\hat \gamma=X_m/(m-X_m)=1.4253$. The resulting Bayes Factors are $B_{222}^{(2)}=0.1539$ which are less than $0.3162$ based on the standard shown in Table \ref{t2} and the posterior probabilities of the null hypothesis and the alternative hypothesis given the data are 
$$
\Pr(H_0^{(2)} :\gamma \le 1 \mid X_{222}=134)=0.1334 \qquad \text{and} \quad \Pr(H_1^{(2)} :\gamma > 1 \mid X_{222}=134)=0.8666,
$$
which are clearly smaller than the assumed prior probability $\pi_0=0.5$ of $H_0^{(2)}$ being true (marked in a red line in Figure \ref{fj1}).

On the other hand, for the hypotheses of Case $3$, the observation process would 'terminate' much earlier at the $14^{th}$-group data point with a total of $m=172$ 'events' of observed side effect of which $X_m=91$ were noted in the treatment group, leading to the estimated relative risk of $\hat \gamma=X_m/(m-X_m)=1.1235$. The resulting Bayes Factor is $B_{172}^{(3)}=0.2869$ which are less than $0.3162$ based on the standard shown in Table \ref{t2} and the posterior probabilities of the null hypothesis and the alternative hypothesis given the data are 
$$
\Pr(H_0^{(3)} :\gamma \le 1 \mid X_{172}=91)=0.2229 \qquad \text{and} \quad \Pr(H_1^{(3)} :\gamma > 1 \mid X_{172}=91)=0.7771,
$$
which are clearly smaller than the assumed prior probability $\pi_0=0.5$ of $H_0^{(3)}$ being true (marked in a red line in Figure \ref{fj1}).

\begin{table}[H]
\begin{center}{\footnotesize
\caption{The values of Bayes factor and posterior probability of $H_0$ for each data point among the 3 different cases with the Jeffrey's prior.}
\label{tj1}
\begin{tabular}{ c c c c c c c c c c }
\hline
Test & $m$ & $X_m$ & $\hat{\gamma}$ & $B_m^{(1)}$ & $\Pr(H_0^{(1)}\mid X_m)$ & $B_m^{(2)}$ & $\Pr(H_0^{(2)}\mid X_m)$ & $B_m^{(3)}$ & $\Pr(H_0^{(3)}\mid X_m)$ \\
\hline
1  & 12  & 1 &0.0909 &0.0348 &0.0337 &34.7839 &0.9721 &997.4399 &0.9990 \\
2   &18   &5 &0.3846 &0.8570 &0.4615 &30.4686 &0.9682 &34.5543 &0.9719\\
3   &24  &11 &0.8462 &5.7079 &0.8509 &16.6892 &0.9435 &1.9239 &0.6580\\
4   &30  &15 &1.0000 &6.9221 &0.8738 &13.8442 &0.9326 &1.0000 &0.5000\\
5   &34  &15 &0.7895 &5.8156 &0.8533 &23.5774 &0.9593 &3.0541 &0.7533\\
6   &40  &17 &0.7391 &5.0778 &0.8355 &29.6152 &0.9673 &4.8323 &0.8285\\
7   &46  &20 &0.7692 &5.7730 &0.8524 &30.6617 &0.9684 &4.3113 &0.8117\\
8   &67  &34 &1.0303 &10.2206 &0.9109 &18.6318 &0.9491 &0.8230 &0.4514\\
9   &78 &39 &1.0000 &11.1045 &0.9174 &22.2090 &0.9569 &1.0000 &0.5000\\
10 &100  &44 &0.7857 &6.1053 &0.8593 &53.0962 &0.9815 &7.6967 &0.8850\\
11 &115  &51 &0.7969 &6.4501 &0.8658 &57.2681 &0.9828 &7.8787 &0.8874\\
12 &135  &63 &0.8750 &10.8056 &0.9153 &49.2625 &0.9801 &3.5590 &0.7807\\
13 &167  &88 &1.1139 &12.7261 &0.9271 &16.8142 &0.9439 &0.3212 &0.2431\\
14 &172  &91 &1.1235 &12.3066 &0.9248 &15.8372 &0.9406 &0.2869 &0.2229\\
15 &190 &107 &1.2892 &3.7840 &0.7910 &3.9445 &0.7978 &0.0424 &0.0407\\
16 &197 &113 &1.3452 &2.0676 &0.6740 &2.1082 &0.6783 &0.0196 &0.0193\\
17 &211 &124 &1.4253 &0.6991 &0.4114 &0.7028 &0.4127 &0.0054 &0.0053\\
18 &218 &130 &1.4773 &0.3160 &0.2401 &0.3167 &0.2405 &0.0022 &0.0022\\
19 &222 &134 &1.5227 &0.1538 &0.1333 &0.1539 &0.1334 &0.0010 &0.0010\\
20 &231 &141 &1.5667 &0.0653 &0.0613 &0.0653 &0.0613 &0.0004 &0.0004\\
21 &240 &148 &1.6087 &0.0266 &0.0259 &0.0266 &0.0259 &0.0001 &0.0001\\
22 &245 &153 &1.6630 &0.0091 &0.0090 &0.0091 &0.0090 &0.0000 &0.0000\\
23 &247 &155 &1.6848 &0.0058 &0.0058 &0.0058 &0.0058 &0.0000 &0.0000\\
24 &251 &157 &1.6702 &0.0067 &0.0067 &0.0067 &0.0067 &0.0000 &0.0000\\
\hline
\end{tabular}
}
\end{center}
\end{table}

In conclusion, under the Jeffrey's prior, Case $1$ would reject and be 'terminated' at the $18^{th}$-group data point, Case $2$ would reject and 'terminated' at the $19^{th}$-group data point, and Case $3$ would reject and 'terminated' at the $14^{th}$-group data point.

\subsection{The modified Bayesian Test}

\begin{figure}[H]
\centering
\begin{subfigure}[b]{0.328\textwidth}
\centering
\includegraphics[width=\textwidth]{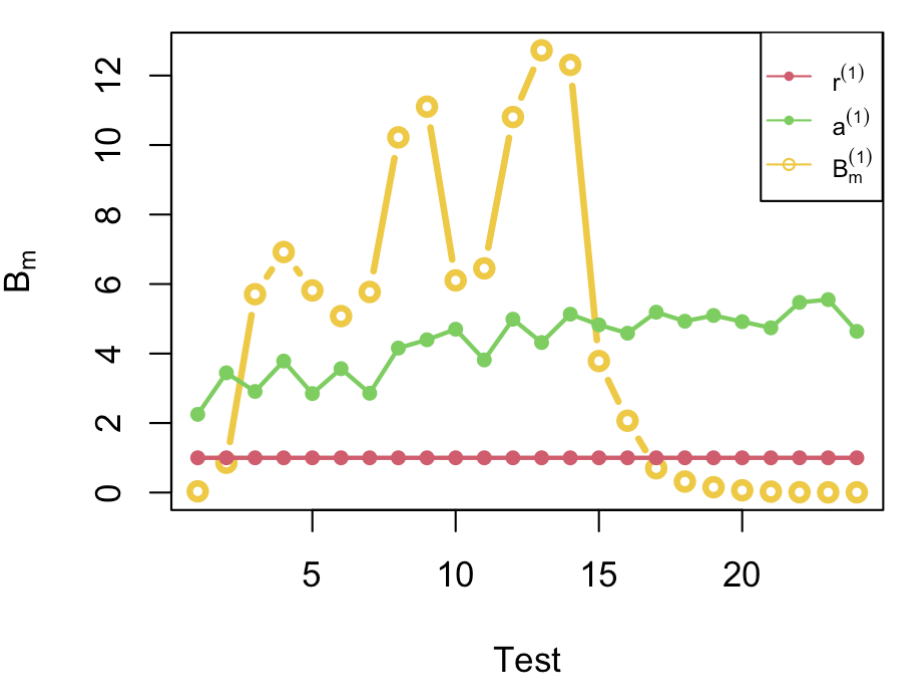}
\caption{Case 1}
\end{subfigure}
\hfill
\begin{subfigure}[b]{0.328\textwidth}
\centering
\includegraphics[width=\textwidth]{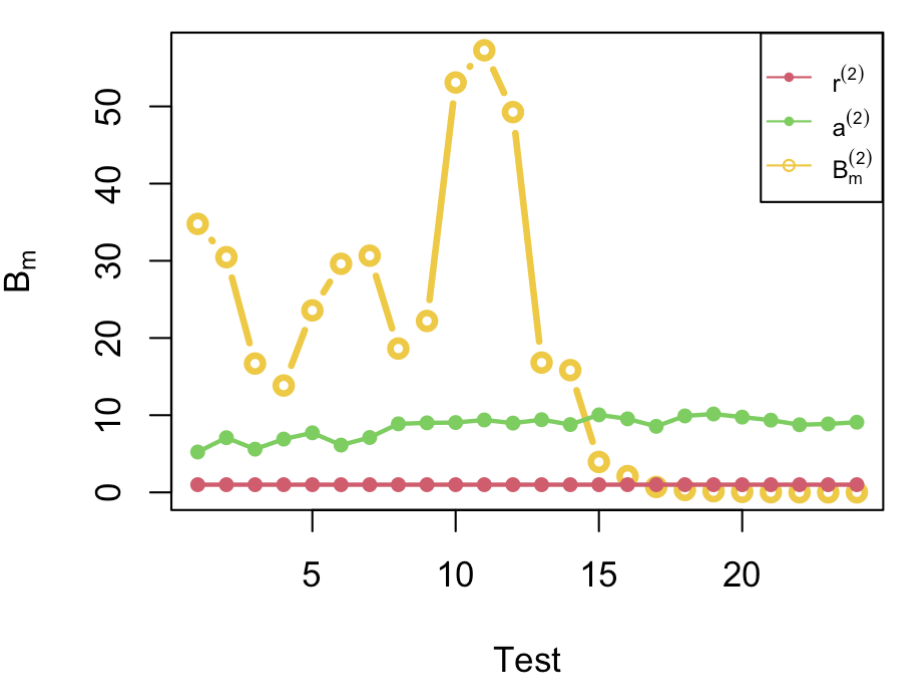}
\caption{Case 2}
\end{subfigure}
\begin{subfigure}[b]{0.328\textwidth}
\centering
\includegraphics[width=\textwidth]{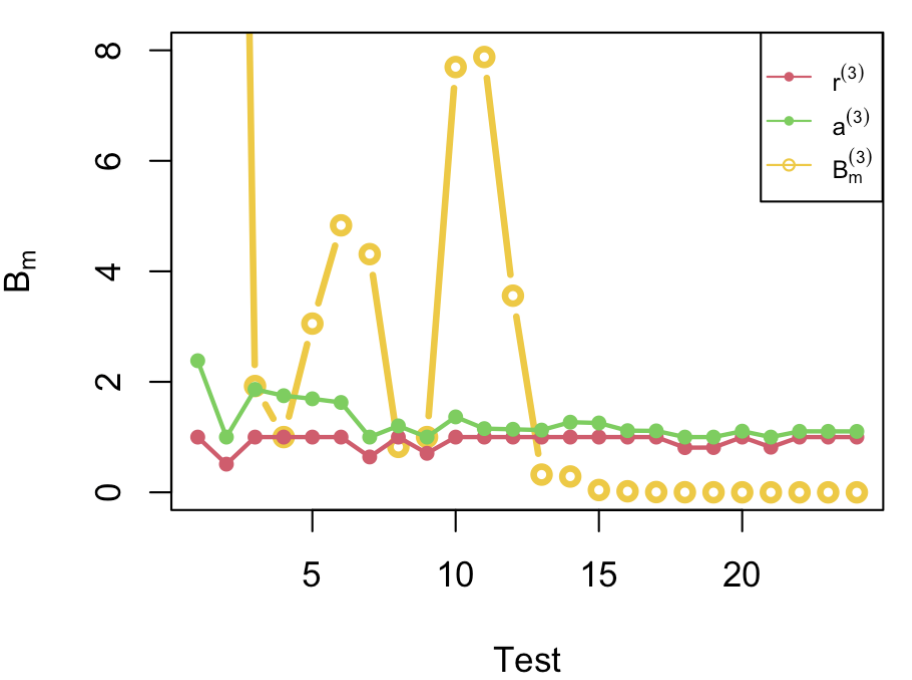}
\caption{Case 3}
\end{subfigure}
\caption{The acceptance and rejection boundaries of $B_m$ for the 24 data points of Table \ref{tj2}}
\label{fj2}
\end{figure}

In the modified Bayesian Sequential test, we obtain the same conclusions that were obtained utilizing the standard Bayesian test for the hypotheses $H_0^{(i)}$, $i=1,2,3$ (see Section \ref{s5.1}). The results corresponding to the Jeffrey's prior distribution are shown in Figure \ref{fj2}, Table \ref{tj2} and Table \ref{tj3}. The calculated conditional Type~I and Type~II error probabilities, $\alpha^*$ and $\beta^*$, corresponding to the 'Reject' or 'Accept' decision (see Table \ref{tj3}).

\begin{table}[H]
\begin{center}{\footnotesize 
\caption{The values of Bayes factor and the boundaries based on it for each data point among different cases with the Jeffrey's prior.}
\label{tj2}
\begin{tabular}{ c c c c c c c c c c c c c }
\hline
Test & $m$ & $X_m$ & $\hat{\gamma}$ & $B_m^{(1)}$ & $r^{(1)}$ &$a^{(1)}$ & $B_m^{(2)}$ &$r^{(2)}$ &$a^{(2)}$ & $B_m^{(3)}$ &$r^{(3)}$ &$a^{(3)}$ \\
\hline
1  & 12  & 1 &0.0909 &0.0348 &1 &2.2505 &34.7839 &1 &5.2308 &997.4399 &1 &2.3824 \\
2   &18   &5 &0.3846 &0.8570 &1 &3.4458 &30.4686 &1 &7.0875 &34.5543 &0.5104 &1\\
3   &24  &11 &0.8462 &5.7079 &1 &2.9086 &16.6892 &1 &5.6013 &1.9239 &1 &1.8636\\
4   &30  &15 &1.0000 &6.9221 &1 &3.7844 &13.8442 &1 &6.9070 &1.0000 &1 &1.7492\\
5   &34  &15 &0.7895 &5.8156 &1 &2.8478 &23.5774 &1 &7.7198 &3.0541 &1 &1.6929\\
6   &40  &17 &0.7391 &5.0778 &1 &3.5652 &29.6152 &1 &6.1286 &4.8323 &1 &1.6267\\
7   &46  &20 &0.7692 &5.7730 &1 &2.8578 &30.6617 &1 &7.1120 &4.3113 &0.6408 &1\\
8   &67  &34 &1.0303 &10.2206 &1 &4.1572 &18.6318 &1 &8.8826 &0.8230 &1 &1.2029\\
9   &78 &39 &1.0000 &11.1045 &1 &4.3958 &22.2090 &1 &9.0068 &1.0000 &0.7064 &1\\
10 &100  &44 &0.7857 &6.1053 &1 &4.7006 &53.0962 &1 &9.0504 &7.6967 &1 &1.3656\\
11 &115  &51 &0.7969 &6.4501 &1 &3.8165 &57.2681 &1 &9.3841 &7.8787 &1 &1.1532\\
12 &135  &63 &0.8750 &10.8056 &1 &4.9882 &49.2625 &1 &8.9751 &3.5590 &1 &1.1411\\
13 &167  &88 &1.1139 &12.7261 &1 &4.3165 &16.8142 &1 &9.4196 &0.3212 &1 &1.1266\\
14 &172  &91 &1.1235 &12.3066 &1 &5.1324 &15.8372 &1 &8.7899 &0.2869 &1 &1.2698\\
15 &190 &107 &1.2892 &3.7840 &1 &4.8263 &3.9445 &1 &10.0463 &0.0424 &1 &1.2555\\
16 &197 &113 &1.3452 &2.0676 &1 &4.5858 &2.1082 &1 &9.5259 &0.0196 &1 &1.1163\\
17 &211 &124 &1.4253 &0.6991 &1 &5.1907 &0.7028 &1 &8.5552 &0.0054 &1 &1.1123\\
18 &218 &130 &1.4773 &0.3160 &1 &4.9304 &0.3167 &1 &9.9074 &0.0022 &0.8092 &1\\
19 &222 &134 &1.5227 &0.1538 &1 &5.0959 &0.1539 &1 &10.1571 &0.0010 &0.8107 &1\\
20 &231 &141 &1.5667 &0.0653 &1 &4.9166 &0.0653 &1 &9.7513 &0.0004 &1 &1.1072\\
21 &240 &148 &1.6087 &0.0266 &1 &4.7401 &0.0266 &1 &9.3589 &0.0001 &0.8171 &1\\
22 &245 &153 &1.6630 &0.0091 &1 &5.4724 &0.0091 &1 &8.7620 &0.0000 &1 &1.1040\\
23 &247 &155 &1.6848 &0.0058 &1 &5.5520 &0.0058 &1 &8.8709 &0.0000 &1 &1.1036\\
24 &251 &157 &1.6702 &0.0067 &1 &4.6391 &0.0067 &1 &9.0885 &0.0000 &1 &1.1027\\
\hline
\end{tabular}
}
\end{center}
\end{table}

\begin{table}[H]
\begin{center}{\scriptsize
\caption{The decision and corresponding conditional $\alpha^*$ and $\beta^*$ for each data point among different cases with the Jeffrey's prior. (Decision: 'R' indicates Reject, 'A' indicates Accept, 'ND' indicates No Decision, 'NA' indicates No Answer.)}
\label{tj3}
\begin{tabular}{ c c c c c c c c c c c }
\hline
Test  & $\hat{\gamma}$ & $B_m^{(1)}$ & Decision &$\alpha^*(\text{or } \beta^*)$ & $B_m^{(2)}$ &Decision &$\alpha^*(\text{or } \beta^*)$ & $B_m^{(3)}$ &Decision &$\alpha^*(\text{or } \beta^*)$ \\
\hline
1   &0.0909 &0.0348 &R &0.0336 &34.7839 &A &0.0286 &997.4399 &A &0.0010 \\
2    &0.3846 &0.8570 &R &0.4615 &30.4686 &A &0.0319 &34.5543 &A &0.0256\\
3    &0.8462 &5.7079 &A &0.1773 &16.6892 &A &0.0612 &1.9239 &A &0.3304\\
4    &1.0000 &6.9221 &A &0.1263 &13.8442 &A &0.0714 &1.0000 &ND &NA\\
5    &0.7895 &5.8156 &A &0.1490 &23.5774 &A &0.0462 &3.0541 &A &0.2449\\
6    &0.7391 &5.0778 &A &0.1773 &29.6152 &A &0.0359 &4.8323 &A &0.1427\\
7   &0.7692 &5.7730 &A &0.1490 &30.6617 &A &0.0319 &4.3113 &A &0.1730\\
8    &1.0303 &10.2206 &A &0.0937 &18.6318 &A &0.0530 &0.8230 &R &0.4515\\
9   &1.0000 &11.1045 &A &0.0937 &22.2090 &A &0.0462 &1.0000 &ND &NA\\
10  &0.7857 &6.1053 &A &0.1490 &53.0962 &A &0.0195 &7.6967 &A &0.0934\\
11  &0.7969 &6.4501 &A &0.1490 &57.2681 &A &0.0179 &7.8787 &A &0.0934\\
12  &0.8750 &10.8056 &A &0.0937 &49.2625 &A &0.0213 &3.5590 &A &0.2071\\
13  &1.1139 &12.7261 &A &0.0730 &16.8142 &A &0.0612 &0.3212 &R &0.2431\\
14  &1.1235 &12.3066 &A &0.0821 &15.8372 &A &0.0612 &0.2869 &R &0.2229\\
15  &1.2892 &3.7840 &ND &NA &3.9445 &ND &NA &0.0424 &R &0.0407\\
16  &1.3452 &2.0676 &ND &NA &2.1082 &ND &NA &0.0196 &R &0.0192\\
17  &1.4253 &0.6991 &R &0.4115 &0.7028 &R &0.4127 &0.0054 &R &0.0054\\
18  &1.4773 &0.3160 &R &0.2401 &0.3167 &R &0.2405 &0.0022 &R &0.0022\\
19  &1.5227 &0.1538 &R &0.1333 &0.1539 &R &0.1334 &0.0010 &R &0.0010\\
20  &1.5667 &0.0653 &R &0.0613 &0.0653 &R &0.0613 &0.0004 &R &0.0004\\
21  &1.6087 &0.0266 &R &0.0259 &0.0266 &R &0.0259 &0.0001 &R &0.0001\\
22  &1.6630 &0.0091 &R &0.0090 &0.0091 &R &0.0090 &0.0000 &R &0.0000\\
23  &1.6848 &0.0058 &R &0.0058 &0.0058 &R &0.0058 &0.0000 &R &0.0000\\
24  &1.6702 &0.0067 &R &0.0067 &0.0067 &R &0.0067 &0.0000 &R &0.0000\\
\hline
\end{tabular}
}
\end{center}
\end{table}

\end{document}